# Sea ice and methane

**Clive Hambler (1)\*, Peter A. Henderson (1,2)**

((1) Department of Zoology, University of Oxford, UK, ORCID 0000-0002-2361-828X (2) PISCES Conservation Ltd, UK, ORCID 0000-0002-7461-1758).   * corresponding author clive.hambler@zoo.ox.ac.uk

**Abstract**

1) The annual cycle of atmospheric methane in southern high latitudes is extremely highly correlated with Antarctic sea ice extent.  2) The annual cycle of atmospheric methane in the Arctic is highly correlated with Antarctic or Arctic plus Antarctic sea ice extent.  3) We propose the global annual cycle of atmospheric methane is largely driven by Antarctic sea ice dynamics, with relatively stronger influence from other fluxes (probably the biota) in the Northern Hemisphere.  4) We propose degassing during sea ice freeze and temperature dependent solubility in the ocean dominate the annual methane cycle.  5) Results provide evidence that carbon cycle pathways, parameters and predictions must be reassessed.

## Introduction

Atmospheric methane is widely considered an important greenhouse gas (Salby; 2012;  IPCC 2013;  Reinhard et al 2020) and the current understanding of its dynamics has shaped climate policies relating to livestock, other agriculture and human diet.  However, many parameters of the methane component of the carbon cycle are not well quantified, including many sources and sinks (Salby 2012; Ciais et al 2013;  Matthews et al 2020).  Emissions in the Northern Hemisphere are believed to dominate (Khalil & Rasmussen 1983;  Semiletov 1999;  Ciais et al 2013).  The prevailing explanations for seasonal variation have been emissions from tundra and lakes, and scavenging by the hydroxyl radical (OH) (Mastepanov et al 2008;  Salby, 2012;  Ciais et al 2013;  Matthews et al 2020).  The regional variation in cycles has been used in validation of Earth system models (He et al 2020).  Here we demonstrate that annual Antarctic sea ice dynamics are highly correlated with annual atmospheric methane dynamics, and suggest that other seasonal sources and sinks are therefore relatively smaller than generally believed.

We hypothesize a major role for sea ice in annual cycle of methane;  the current paradigm does not include a substantive involvement of sea ice (Ciais et al 2013;  He et al 2020) although it is included with highly limited mechanistic detail in some Earth system models (Reinhard et al 2020).

Air-sea fluxes of carbon compounds are sparsely sampled and partially quantified, particularly in sea ice zones (Ciais et al 2013; Takahashi et al 1993, 2009;  Rosso et al 2017;  Tison et al 2017;  Vancoppenolle & Tedesco 2017;  Geilfus et al 2018;  Gray et al 2018;  Resplandy et al 2018;  Bushinsky et al 2019;  Francey et al 2019;  MOSAiC 2019;  Ouyang et al 2020;  Reinhard et al 2020).

We address these questions regarding the annual (seasonal) cycle of methane and regional sea ice extent:

1)  How do Antarctic sea ice dynamics relate to atmospheric methane dynamics?

2)  How do Arctic sea ice dynamics relate to atmospheric methane dynamics?

3)  How do 'global' (Arctic plus Antarctic) sea ice dynamics relate to 'global' atmospheric methane dynamics recorded at Mauna Loa?

**Methods**

*Data*

We use the datasets in Table 1.

**Table 1**  Data sources

| Variable | Source |
|---|---|
| Atmospheric methane ($CH_4$)<br><br>NOAA Global Monitoring Laboratory<br>Earth System Research Laboratories<br>data and site name abbreviations | https://www.esrl.noaa.gov/gmd/dv/data/<br>NOAA GML Carbon Cycle Cooperative Global Air Sampling<br>Network, 1983-2019, Version: 2020-07,<br>https://doi.org/10.15138/VNCZ-M766<br><br>ftp://aftp.cmdl.noaa.gov/data/trace_gases/ch4/flask<br><br>Accessed 1 August 2020<br><br>Dlugokencky et al 2020a |
| Atmospheric carbon dioxide ($CO_2$)<br><br>NOAA Global Monitoring Laboratory<br>Earth System Research Laboratories | https://www.esrl.noaa.gov/gmd/dv/data/<br>NOAA GML Carbon Cycle Cooperative Global Air Sampling<br>Network, 1968-2019, Version: 2020-07,<br>https://doi.org/10.15138/wkgj-f215<br><br>ftp://aftp.cmdl.noaa.gov/data/trace_gases/co2/flask<br><br>Accessed 1 August 2020<br><br>Dlugokencky et al 2020b |
| Sea ice extent<br><br>(NSIDC) | https://nsidc.org/data/seaice_index/archives<br>Sea Ice Index Version 3<br>ftp://sidads.colorado.edu/DATASETS/NOAA/G02135/<br>(Fetterer et al 2017)<br><br>'North' (= 'Arctic'):<br>ftp://sidads.colorado.edu//DATASETS/NOAA/G02135/north/m<br>onthly/data/ at sidads.colorado.edu<br><br>'South' (= 'Antarctic'):<br>ftp://sidads.colorado.edu/DATASETS/NOAA/G02135/south/m<br>onthly/data/<br>Files in form:  S_01_extent_v3.0.csv<br><br>Accessed 26 February 2020 |

In time series, causal variables must be correlated with response variables, with or without time lags.  Short lags and strong correlations between variables make detection of potential causation more straightforward.  We use cross-correlations to test if sea ice change correlates with methane change, bearing in mind difficulties in ascribing causality in time series (Faes et al 2017).  We examine the first derivative of atmospheric $CH_4$ level, which we term the 'methane rate'.  We compare this with the first derivative of sea ice extent, which we term the 'sea ice rate' and the first derivative of atmospheric carbon dioxide level, which we term the carbon dioxide rate.  In the annual cycle, if sea ice change is largely causal of atmospheric methane change the methane rate must be synchronous with or lag the sea ice rate, and a very strong correlation would make this more plausible.  Similarities between methane and carbon dioxide rates might help elucidate mechanisms.

Global monthly extent of sea ice was calculated using the sum of Arctic and Antarctic sea ice monthly extent.  For months with missing daily data, averages were taken for the months with data.

Rates of change of variables were approximated as follows:  changes in month 2 were taken as the value in month 2 minus the value in month 1, and are plotted in month 2.

We compare methane rates for six sites selected from those where there are long term methane records from flasks:

a)  The South Pole (elevation 2810 metres above sea level), a site used in IPCC 2013.

b)  Palmer Station, Antarctic Peninsula (10 masl), a coastal Antarctic site with an estimated local atmospheric measurement footprint (Le Quéré et al 2007).

c)  Barrow, Point Barrow, Alaska (11 masl), a northern site often compared with Mauna Loa in publications on carbon dioxide (e.g. Graven et al 2013).

d)  Alert, Canada (200 masl), being the site nearest the North Pole and used in IPCC (2103).

e)  Summit, Greenland (3210 masl), being the high altitude recording site closest to the North Pole.

f)  Mauna Loa, Hawaii, USA, (3397 masl), the widely used standard site for measuring global atmospheric $CO_2$ and methane trends (Ciais et al 2013).

Examples of plots of methane rates from other recording sites with fairly continuous recording using flasks are presented in Appendix 1;  these sites are chosen from the NOAA Global Monitoring Laboratory Earth System Research Laboratories network and include most extra-tropical Southern Hemisphere sites.  No statistical analyses were performed on these data, which should be treated as unrepresentative and preliminary but suffice to illustrate there is regional diversity of methane dynamics.

The earliest start date used for any analysis was January 2006, after which the sea ice extent record from NSIDC is continuous.  Monthly rates are thus available from February 2006.  The latest date used was December 2019, since this is the last date for which monthly methane flask levels were available from NOAA for our selected sites at the time of analysis (August 2020).  Monthly flask measurements were used because these are available in consistent format for many recording stations (Dlugokencky et al 2020).

*Statistical analyses*

The extremely high visual similarity of many of the time series for sea ice rate and methane rate (Figs. 1 - 8 and Appendix 1) makes formal statistical analysis redundant in many cases, particularly in the Southern Hemisphere.  Nevertheless, a few illustrative analyses are performed on selected time series.

All statistical analyses used the R platform. The degree of cross-correlation at lags of up to ± 12 months between time series was calculated using the sample cross-correlation function ccf.  The statistical significance of the correlation at the lag giving the highest cross correlation was obtained using the rcorr function in the Hmisc package. In cases where there is a similar positive or negative lag, the highest bar in the ACF plot is used to deduce the most likely sign of the lag or lead.

We suggest lags longer than a year are unlikely, given that these are not evident in carbon dioxide and sea ice dynamics (Hambler & Henderson, submitted).

**Results**

Results and statistical analyses are presented in Table 2, Figs. 1 - 8 and Appendices.

**Table 2**  Cross-correlations between sea ice extent rates and $CH_4$ rates.

| Variables | Cross-correlation coefficient $r$ and lag at which correlation strongest | p value | Fig. number |
|---|---|---|---|
| South Pole $CH_4$ rate *vs* Antarctic sea ice extent rate | $r = 0.96$ at zero lag | $p < 0.001$ | 1 |
| Palmer Station $CH_4$ rate *vs* Antarctic sea ice extent rate | $r = 0.95$ at zero lag | $p < 0.001$ | 2 |
| Mauna Loa $CH_4$ rate *vs* Antarctic sea ice extent rate | $r = 0.58$ when methane lags ice by 5 months | $p < 0.001$ | 3 |
| Manua Loa $CH_4$ rate *vs* 'Arctic plus Antarctic' sea ice extent rate | $r = 0.63$ when methane lags ice by 6 months;  6 month lead is very similar | $p < 0.001$ | 4 |
| Summit, Greenland $CH_4$ rate *vs* Antarctic sea ice extent rate | $r = 0.82$ when methane lags ice by 5 months | $p < 0.001$ | 5 |
| Alert $CH_4$ rate *vs* Antarctic sea ice extent rate | $r = 0.84$ when methane lags ice by 5 months | $p < 0.001$ | 6 |
| Barrow $CH_4$ rate *vs* Arctic sea ice extent rate | $r = 0.67$ when methane leads ice by 2 months | $p < 0.001$ | 7 |
| Barrow $CH_4$ rate *vs* 'Arctic plus Antarctic' sea ice extent rate | $r = 0.49$ when methane lags ice by 4 months | $p < 0.001$ | 8 |

**Fig. 1**  Monthly South Pole methane rate *vs.* monthly Antarctic sea ice extent rate.  *r* = 0.96 at lag = zero months; p < 0.001

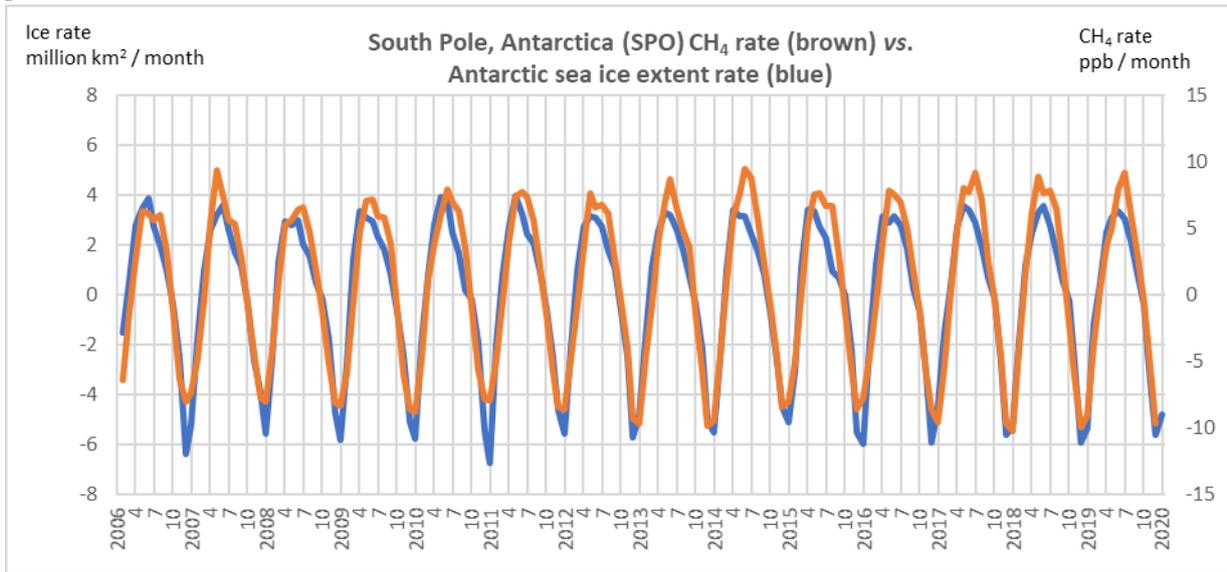

**Fig. 2**  Monthly Palmer Station (Antarctica) methane rate *vs.* monthly Antarctic sea ice extent rate.  *r* = 0.95 at lag = zero months;  p < 0.001

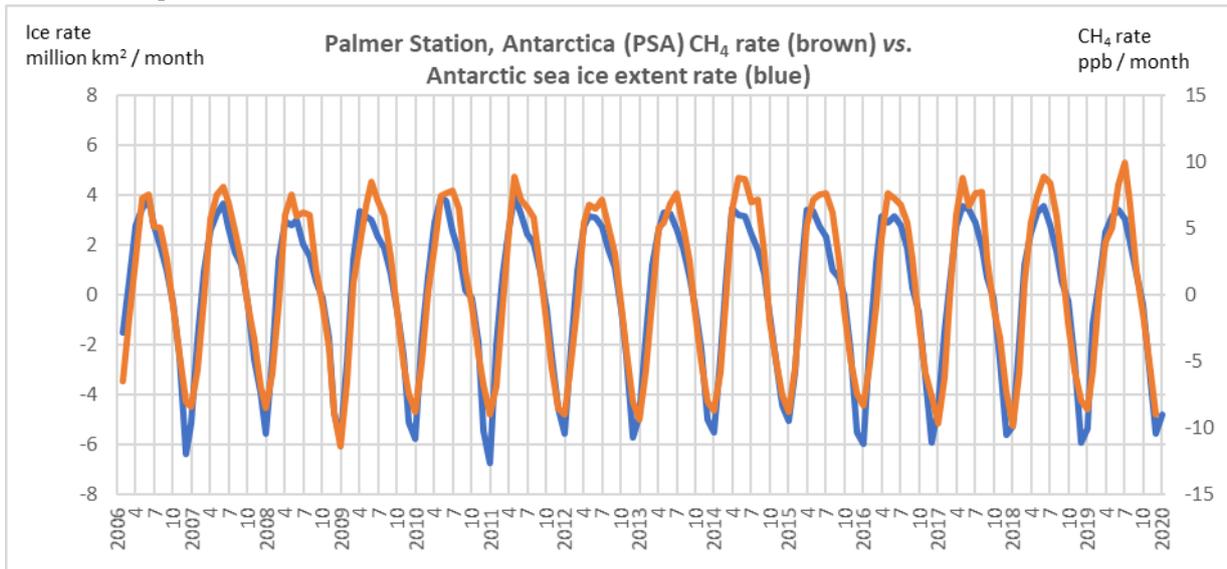

**Fig. 3** Monthly Mauna Loa (Hawaii) methane rate *vs.* monthly Antarctic sea ice extent rate. $r = 0.58$ at lag = 5 months; $p < 0.001$

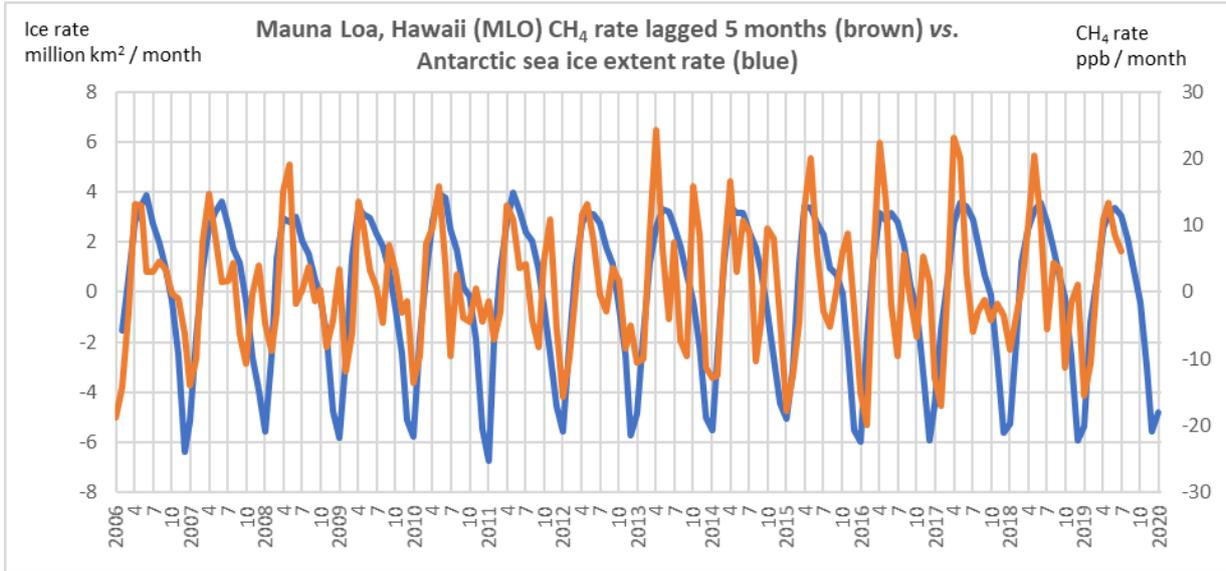

**Fig. 4** Monthly Mauna Loa (Hawaii) methane rate *vs.* monthly Arctic plus Antarctic sea ice extent rate. $r = 0.63$ at lag = 6 months; $p < 0.001$

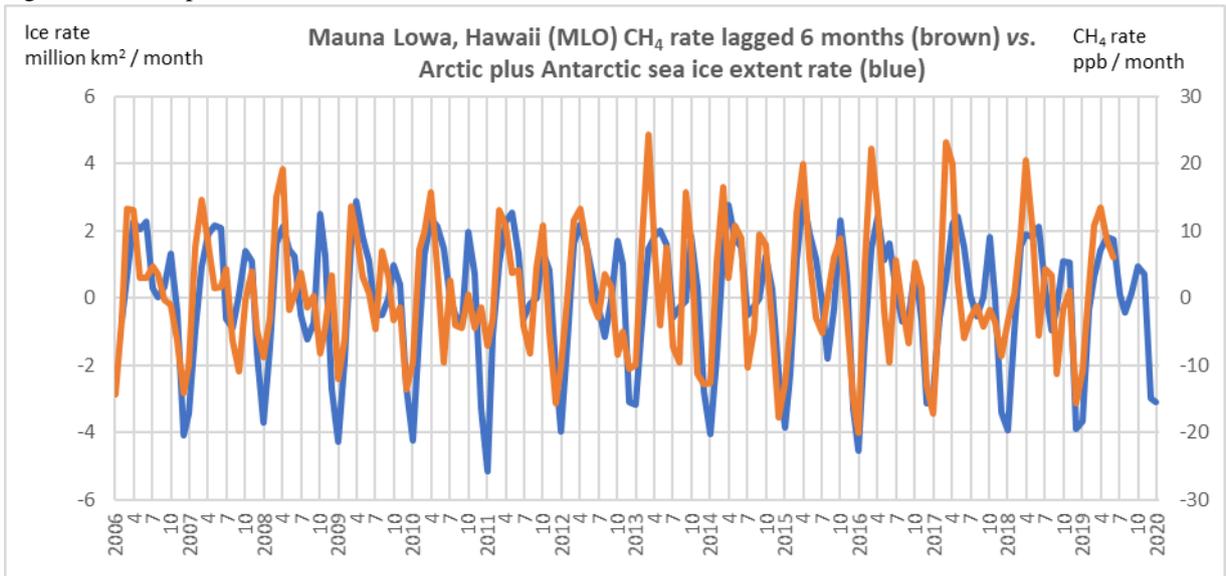

**Fig. 5**  Monthly Summit, Greenland methane rate *vs.* monthly Antarctic sea ice extent rate.  $r = 0.82$ at lag = 5 months;  $p < 0.001$

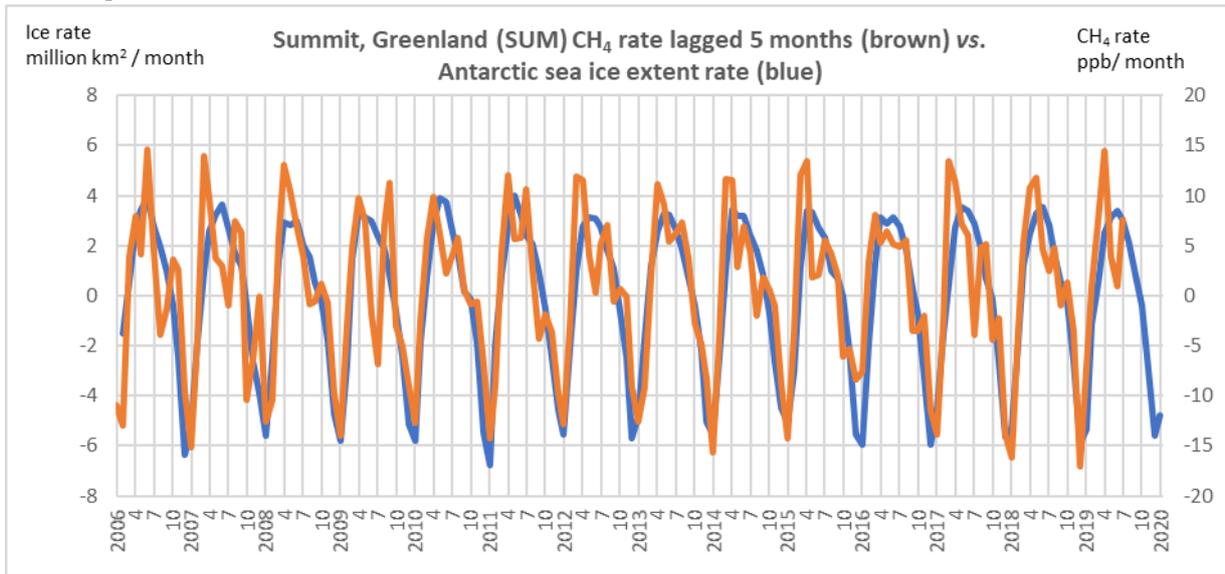

**Fig. 6**  Monthly Alert, Canada, methane rate *vs.* monthly Antarctic sea ice extent rate.  $r = 0.84$ at lag = 5 months; $p < 0.001$

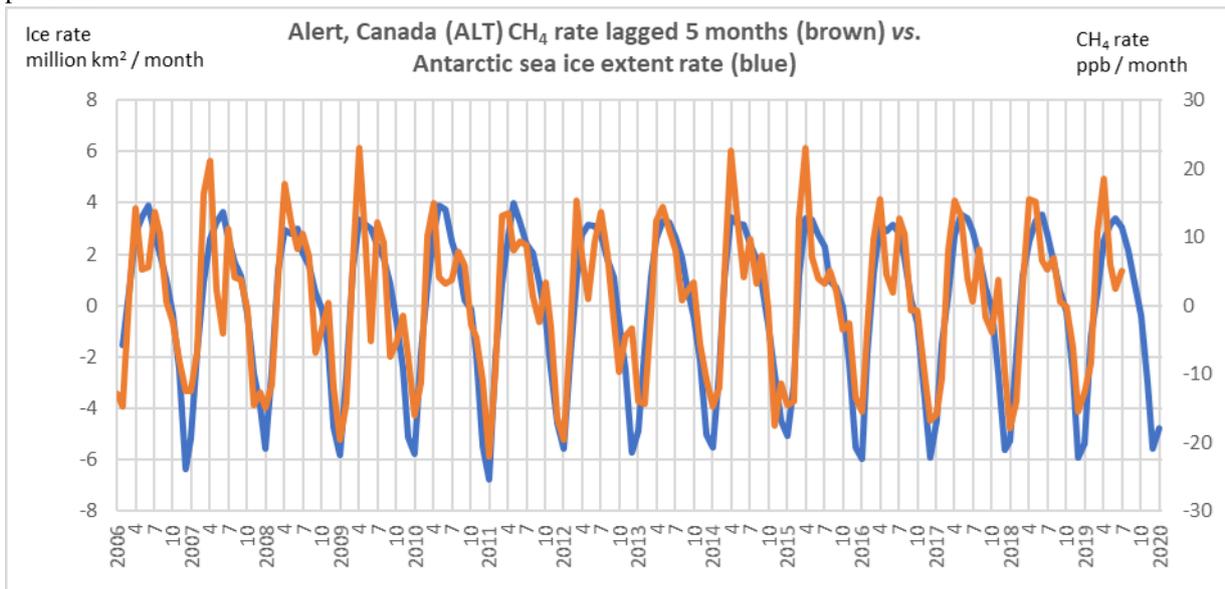

**Fig. 7**  Monthly methane rate Barrow, Alaska, *vs.* monthly Arctic sea ice extent rate.  $r = 0.67$ when methane rate leads sea ice rate by 2 months;  $p < 0.001$

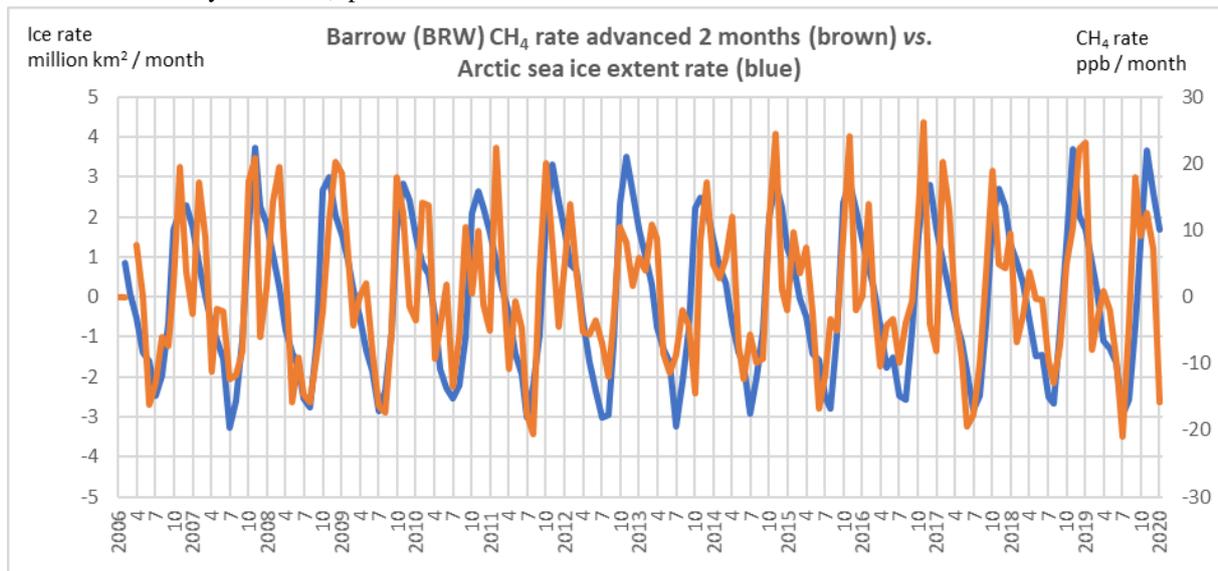

**Fig. 8**  Monthly methane rate Barrow, Alaska, *vs.* monthly Arctic plus Antarctic sea ice extent rate.  $r = 0.49$ at lag = 4 months;  $p < 0.001$

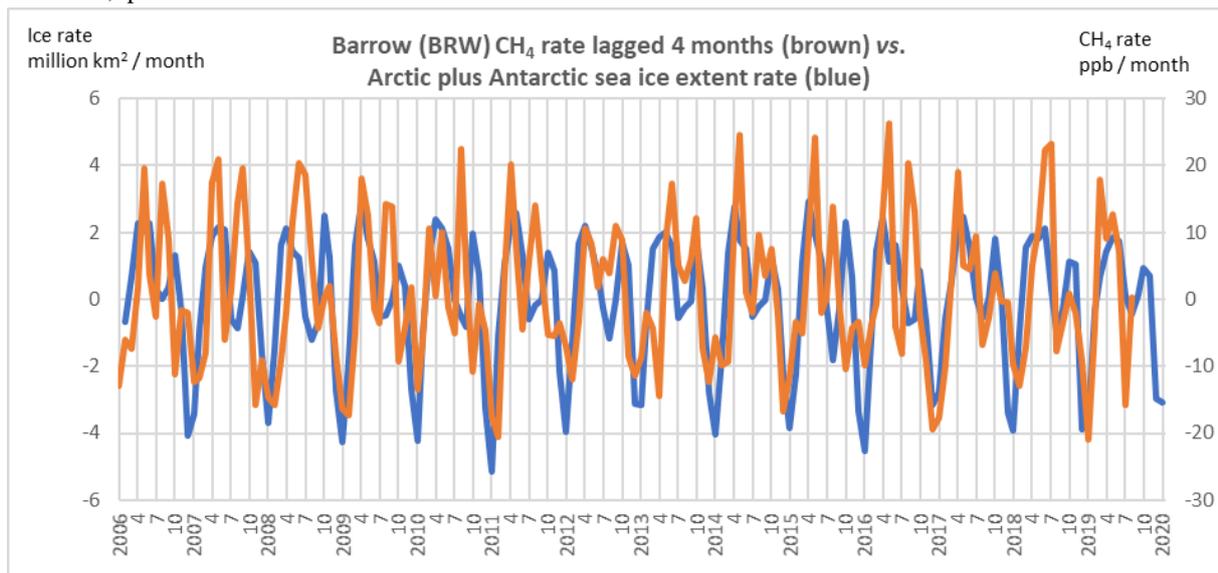

## Discussion

The annual dynamics of methane in the Antarctic region is statistically almost entirely explicable by the annual dynamics of Antarctic sea ice (Table 2;  Figs. 1, 2 and Appendix 1).  The monthly rates of sea ice and methane are highly synchronous:  there is no lag of one month or more at the South Pole and Palmer Station.  Visual inspection of sites in Appendix 1 suggests Halley Station, Syowa Station, Drake Passage, Usuaia (Argentina), Tasmania and New Zealand have no lag between sea ice rate and methane rate.  Indeed, visual inspection (Appendix 1) shows all recording sites in the Southern Hemisphere south of the Tropic of Capricorn (23.4° S) have a clear Antarctic sea ice type phenology (hereafter 'ANP') in the methane rate.  Some sites nearer the equator also have what may be a residual ANP with a visual similarity at a lag of about 3 - 6 months behind the sea ice rate, but in others an ANP is not clear.

Whilst statistical analysis and visual inspection of some sites suggest it is equally likely the methane rate lags or leads the ice rate, in some cases a lag or lead is evident;  in several cases positive and negative lags are nearly identical.  High similarity of the annual cycles across years means longer timeseries might be needed to statistically identify whether the sign of any lag between sea ice rate and methane rate is positive or negative.  However, we consider the simplest explanation (consistent with visually identified lags at many sites in Appendix 1) is that lags are introduced as methane moves through the atmosphere from the Antarctic.

Some equatorial and Northern Hemisphere sites also have highly statistically significant similarities between methane rates and Antarctic sea ice rates if the methane rate is lagged 5 - 7 months (Figs. 3, 5, 6 and Appendix 1).  A high-altitude Arctic site, Summit (72.6° N) has highly statistically significant similarity to Antarctic sea ice, lagged by 5 months (Fig. 5), whilst its methane rate leads Arctic sea ice by about 1 month and hence Arctic sea ice cannot be a dominant cause there within a year.  Sites such as Tudor Hill, Bermuda (32.3° N) and Midway (28.2° N) have high visual similarity to both Antarctic and Arctic sea ice rates (Appendix 1) but with methane lagging the Antarctic rate and leading the Arctic rate, suggesting Antarctic influence could dominate in some sites north of the Tropic of Cancer (23.4° N).  Our selected sites in the Northern Hemisphere have weaker correlations with sea ice rates than do our Southern Hemisphere sites, suggesting a relatively greater role for other likely drivers such as tundra and lakes in the Northern Hemisphere (as in the current paradigm).  Such variables may contribute major emission peaks and influence the complexity of the phenology.  Arctic sea ice rates lag methane rates at Barrow (Fig. 7), Alert (Appendix 1) and Summit - showing they cannot be causal of methane dynamics at these sites on this timescale.

Any Arctic sea ice contribution is stronger in northern sites and is very weak in the Southern Hemisphere.  High latitude sites in the Southern Hemisphere including the South Pole have very strong visual similarity of methane rates to Arctic sea ice rates lagged by about 6 months (Appendix 1), but we suggest this reflects the very similar annual phenology of Arctic and Antarctic sea ice.  Some sites have high statistical (Figs. 4, 8) or visual (Appendix 1) similarity between methane dynamics and the dynamics of the of global sea ice ('Arctic plus Antarctic').

The sites available and used here for methane records are not random, numerous or spatially representative of the globe, although the high-altitude site of Mauna Loa is often taken as globally representative (Ciais et al 2013).  The Mauna Loa methane rate is strongly correlated with the lagged Antarctic sea ice rate (Fig. 3) but even more strongly correlated to the lagged global sea ice rate (Fig. 4), whilst it leads the Arctic sea ice rate by about 1 month (Appendix 1).  The record at Mauna Loa suggests the combined effect of Arctic plus Antarctic sea ice dynamics could dominate the global average methane cycle, but with statistical room for other drivers.

It is evident that atmospheric methane dynamics have similarities across very substantial areas of the globe - particularly if potential atmospheric transport lags of a few months are considered - and may have very few major drivers globally.  However, as illustrated in Appendix 1, some recording sites have a much less clear or regular phenological pattern (such as Mt Waliguan and Ketura) or have very high fluxes (such as Hegyhátsál and the Southern Great Plains) and presumably these reflect strong local sources and / or sinks of methane.  Whilst a systematic comparison of methane recording sites against hemispheric and global sea ice rates would be revealing of widespread patterns, it would still be biased by the limited recording station locations.

We suggest a causal mechanism will be identified in the Antarctic sea ice dynamics which imprints a pattern in methane rates over a very wide area and which attenuates towards the North Pole.  The sea ice zone includes "potentially important" processes in the methane cycle including "intense" emissions (Vancoppenolle et al 2013).  There are "significant" sources of methane in the Arctic Ocean where the sea ice zone emits methane at least locally in some seasons (Vancoppenolle et al 2013), consistent with the twin or multiple peaks of positive methane flux at many recording sites (Figs. 3 - 8 and Appendix 1).

Methane rates are visually synchronous with carbon dioxide rates at Palmer Station and lead carbon dioxide rates by about 1 month at Mauna Loa (Appendix 2). We have previously detected extremely strong correlations between sea ice rates and carbon dioxide rates (Hambler & Henderson, submitted; Appendix 2). We consider it parsimonious to ascribe some features of the annual dynamics of both methane and carbon dioxide to the same basic physics. We suggest the simplest mechanism by which sea ice might drive atmospheric methane would be degassing during the sea water freeze (Nomura et al 2006; Vancoppenolle et al 2013), and dissolution in the cold water during sea ice melt (Wiesenburg & Guinasso 1979; Papadimitriou et al 2004). We anticipate that with suitably extensive, year-round sampling, high levels of dissolved methane will be found in Antarctic waters for much of the year due to upwelling (Talley 2013; Rosso et al 2017). We predict this provides a large methane efflux to the atmosphere on freeze and that a brief, large drawdown of methane occurs into cold undersaturated Antarctic water on sea ice melt before background outgassing resumes.

Other possible drivers of the annual methane cycle might involve the observed changes in photosynthesis, methanogenic bacteria, nutrient availability, anaerobic respiration and marine upwelling during the sea ice freeze-thaw cycle (Vancoppenolle et al 2013), although freezing might be expected to reduce biological activity and upwelling (inconsistent with our results). Given the strength of correlations we find in southern high latitudes, biological processes would have to be remarkably closely and consistently linked to sea ice extent - and from our ecological experience we doubt the biotic response to temperature could be so consistent on these timescales. If sea ice is not a dominant factor driving the annual dynamics of methane, it should be possible to find a stronger correlation with another biotic or abiotic variable and such correlations should be sought. However, given the extremely high correlations we have found we suggest the most productive (and cheapest) approach would be to explore possible sea ice mechanisms first. Large-scale manipulation experiments in the field and laboratory could confirm causality.

**Conclusions**

The dynamics of methane in the atmosphere has not previously been considered a likely consequence of sea ice dynamics, presumably due to confidence in some parameter values and models despite a dearth of recording in high latitudes. We argue sampling and understanding of this component of the carbon cycle is surprisingly poor.

Monthly sea ice rates are forced by solar insolation (temperature) as are emissions from tundra and lakes (Matthews et al 2020). If the temperature dependent dynamics of sea ice drive the annual methane cycle then there is likely to be temperature dependence in inter-annual variation in atmospheric methane, which is consistent with long term data (Salby 2012) and Earth system modelling (He et al 2020). Interannual variation in atmospheric methane has been ascribed with "high confidence" mainly to climatic influence on wetlands (Ciais et al 2013) which merit further observation (Matthews et al 2020). However, sea ice is involved in injection of dissolved carbon compounds to depth (Rysgaard et al 2007; Moreau et al 2016; Tison et al 2017; Ahmed et al 2019) and in air-sea gas exchange (Loose et al 2009; Vancoppenolle 2013; Brown et al 2015; Nomura et al 2018). We argue that Antarctic sea ice has been overlooked and is unlikely to have a neutral annual effect. We also argue the relative contributions of sea ice and humanity to the net annual methane flux require reassessment.

The uncertainties surrounding methane fluxes have implications for many climate models (IPCC 2013; Vancoppenolle & Tedesco 2017) and for all Earth system models (which couple biogeochemical cycles and climate, Notz & Bitz 2017; Hausfather 2019; He et al 2020; Reinhard et al 2020). There are also implications for policies which attempt to manipulate global methane dynamics.

## Acknowledgements

Datasets in the tables are the work of numerous very dedicated researchers including NOAA and sources they acknowledge, and NSIDC. We thank all sources listed or indicated in the metadata on the hosting websites and in Table 1.

## Declarations

**Funding**  None.

**Conflict of interest / Competing interest** The authors declare they have no conflict of interest / competing interests.

**Availability of data and material**  Data are available from the online providers indicated in the Methods.

**Code availability**  R code can be provided upon reasonable request.

## Open Access

## References

Ahmed, M., Else, B., Burgers, T. & Papakyriakou, T. (2019) Variability of surface water $p$CO$_2$ in the Canadian Arctic Archipelago from 2010 to 2016. *Journal of Geophysical Research: Oceans,* **124,** 1876-1896.

Brown, K.A., Miller, L.A., Mundy, C.J., Papakyriakou, T., Francois, R., Gosselin, M., Carnat, G., Swystun, K. & Tortell, P.D. (2015) Inorganic carbon system dynamics in landfast Arctic sea ice during the early-melt period. *Journal of Geophysical Research: Oceans,* **120,** 3542-3566.

Bushinsky, S.M., Landschützer, P., Rödenbeck, C., Gray, A.R., Baker, D., Mazloff, M.R., Resplandy, L., Johnson, K.S. & Sarmiento, J.L. (2019) Reassessing Southern Ocean air sea CO$_2$ flux estimates with the addition of biogeochemical float observations. *Global Biogeochemical Cycles,* **33,** 1370-1388.

Ciais, P., Sabine, C., Bala, G., Bopp, L., Brovkin, V., Canadell, J., Chhabra, A., Defries, R., Galloway, J., Heimann, M., Jones, C., Le Quéré, C., Myneni, R.B., Piao, S. & Thornton, P. (2013) Carbon and other biogeochemical cycles. In: *Climate change 2013: the physical science basis. Contribution of Working Group I to the Fifth Assessment Report of the Intergovernmental Panel on Climate Change* (ed. by T. Stocker, D. Qin, G.-K. Plattner, M. Tignor, S. Allen, J. Boschung, A. Nauels, Y. Xia, V. Bex & P. Midgley), pp. 465-570. Cambridge University Press, Cambridge, United Kingdom.

Dlugokencky, E.J., Crotwell, A.M., Mund, J.W., Crotwell, M.J. & Thoning K.W. (2020a) Atmospheric methane dry air mole fractions from the NOAA GML Carbon Cycle Cooperative Global Air Sampling Network, 1983-2019, Version: 2020-07, https://doi.org/10.15138/VNCZ-M766. Accessed 1 August 2020.

Dlugokencky, E.J., Mund, J.W., Crotwell, A.M., Crotwell M.J. & K.W. Thoning, K.W. (2020b) Atmospheric carbon dioxide dry air mole fractions from the NOAA GML Carbon Cycle Cooperative Global Air Sampling Network, 1968-2019, Version: 2020-07, https://doi.org/10.15138/wkgj-f215. Accessed 1 August 2020.

Faes, L., Nollo, G., Stramaglia, S. & Marinazzo, D. (2017) Multiscale granger causality. *Physical Review E,* **96,** 042150.

Fetterer, F., Knowles, K., Meier, W.N., Savoie, M. & Wn, W. (2017) Updated daily. Sea Ice Index, Version 3. monthly North and South. Boulder, Colorado USA. NSIDC: National Snow and Ice Data Center. https://doi.org/10.7265/N5K072F8. Accessed 26 February 2020.

Francey, R.J., Frederiksen, J.S., Steele, L.P. & Langenfelds, R.L. (2019) Variability in a four-network composite of atmospheric CO$_2$ differences between three primary baseline sites. *Atmospheric Chemistry and Physics,* **19,** 14741-14754.

Geilfus, N.-X., Pind, M., Else, B., Galley, R., Miller, L., Thomas, H., Gosselin, M., Rysgaard, S., Wang, F. & Papakyriakou, T. (2018) Spatial and temporal variability of seawater $p$CO$_2$ within the Canadian Arctic Archipelago and Baffin Bay during the summer and autumn 2011. *Continental Shelf Research,* **156,** 1-10.

Graven, H., Keeling, R., Piper, S., Patra, P., Stephens, B., Wofsy, S., Welp, L., Sweeney, C., Tans, P. & Kelley, J. (2013) Enhanced seasonal exchange of CO$_2$ by northern ecosystems since 1960. *Science,* **341,** 1085-1089.

Gray, A.R., Johnson, K.S., Bushinsky, S.M., Riser, S.C., Russell, J.L., Talley, L.D., Wanninkhof, R., Williams, N.L. & Sarmiento, J.L. (2018) Autonomous biogeochemical floats detect significant carbon dioxide outgassing in the high latitude Southern Ocean. *Geophysical Research Letters,* **45,** 9049-9057.

Hambler, C. & Henderson, P.A. (submitted) Sea ice and carbon dioxide. arXiv submission identifier submit3132691. Submitted 16 April 2020: under appeal.

Hausfather, Z. (2019) CMIP6: the next generation of climate models explained. CarbonBrief. https://www.carbonbrief.org/cmip6-the-next-generation-of-climate-models-explained. Accessed 12 February 2020.

He, J., Naik, V., Horowitz, L.W., Dlugokencky, E. & Thoning, K. (2020) Investigation of the global methane budget over 1980–2017 using GFDL-AM4.1 *Atmospheric Chemistry and Physics,* **20,** 805–827.

IPCC (2013) *Climate Change 2013: The physical science basis. Contribution of Working Group I to the Fifth Assessment Report of the Intergovernmental Panel on Climate Change* (ed. by T. Stocker, D. Qin, G.-K. Plattner, M. Tignor, S. Allen, J. Boschung, A. Nauels, Y. Xia, V. Bex & P. Midgley), Cambridge University Press, Cambridge, United Kingdom.

Khalil, M.A.K & Rasmussen, R.A (1983) Sources, sinks, and seasonal cycles of atmospheric methane. *Journal of Geophysical Research: Oceans,* **88,** 5131-5144.

Le Quéré, C., Rödenbeck, C., Buitenhuis, E.T., Conway, T.J., Langenfelds, R., Gomez, A., Labuschagne, C., Ramonet, M., Nakazawa, T. & Metzl, N. (2007) Saturation of the Southern Ocean CO$_2$ sink due to recent climate change. *Science,* **316,** 1735-1738.

Loose, B., Mcgillis, W., Schlosser, P., Perovich, D. & Takahashi, T. (2009) Effects of freezing, growth, and ice cover on gas transport processes in laboratory seawater experiments. *Geophysical Research Letters,* **36,** L05603.

Mastepanov, M., Sigsgaard, C., Dlugokencky, E.J., Houweling, S., Ström, L., Tamstorf, M.P. & Torben R. Christensen, T.R. (2008) Large tundra methane burst during onset of freezing. *Nature,* **456,** 628–630.

Matthews, E., Johnson, M.S., Genovese, V., Du, J., & Bastviken, D. (2020) Methane emission from high latitude lakes: methane-centric lake classification and satellite-driven annual cycle of emissions *Scientific Reports,* **10,** 12465.

Moreau, S., Vancoppenolle, M., Bopp, L., Aumont, O., Madec, G., Delille, B., Tison, J.-L., Barriat, P.-Y. & Goosse, H. (2016) Assessment of the sea-ice carbon pump: Insights from a three-dimensional ocean-sea-ice biogeochemical model (NEMO-LIM-PISCES). *Elementa: Science of the Anthropocene,* **4,** 000122.

MOSAiC (2019) The key to the Arctic puzzle. from https://www.mosaic-expedition.org/science/arctic-climate/. Accessed 29 October 2019.

Nomura, D., Granskog, M.A., Fransson, A., Chierici, M., Silyakova, A., Ohshima, K.I., Cohen, L., Delille, B., Hudson, S.R. & Dieckmann, G.S. (2018) CO$_2$ flux over young and snow-covered Arctic pack ice in winter and spring. *Biogeosciences,* **15,** 3331-3343.

Notz, D. & Bitz, C.M. (2017) Sea Ice in Earth System Models. In: *Sea ice,* 3rd edn. (ed. by D. Thomas), pp. 304-325. Wiley, New Jersey.

Ouyang, Z., Qi, D., Chen, L., Takahashi, T., Zhong, W., DeGrandpre, M.D., Chen, B., Gao, Z., Nishino, S., Murata, A., Sun, H., Robbins, L.L, Jin, M. & Cai, W.-J. (2020) Sea-ice loss amplifies summertime decadal CO$_2$ increase in the western Arctic Ocean, *Nature Climate Change.* https://doi.org/10.1038/s41558-020-0784-2 Accessed 18 June 2020.

Papadimitriou, S., Kennedy, H., Kattner, G., Dieckmann, G. & Thomas, D. (2004) Experimental evidence for carbonate precipitation and CO$_2$ degassing during sea ice formation. *Geochimica et Cosmochimica Acta,* **68,** 1749-1761.

Reinhard, C.T., Olson, S.L., Kirtland Turner, S., Pälike, C., Kanzaki, Y. & Ridgwell, A. (2020) Oceanic and atmospheric methane cycling in the cGENIE Earth system model. https://arxiv.org/ftp/arxiv/papers/2007/2007.15053.pdf. Accessed 1 August 2020.

Resplandy, L., Keeling, R., Rödenbeck, C., Stephens, B., Khatiwala, S., Rodgers, K., Long, M., Bopp, L. & Tans, P. (2018) Revision of global carbon fluxes based on a reassessment of oceanic and riverine carbon transport. *Nature Geoscience,* **11,** 504-509.

Rosso, I., Mazloff, M.R., Verdy, A. & Talley, L.D. (2017) Space and time variability of the Southern Ocean carbon budget. *Journal of Geophysical Research: Oceans,* **122,** 7407-7432.

Rysgaard, S., Glud, R.N., Sejr, M., Bendtsen, J. & Christensen, P. (2007) Inorganic carbon transport during sea ice growth and decay: A carbon pump in polar seas. *Journal of Geophysical Research: Oceans,* **112,** C03016.

Salby, M. (2012) *Physics of the atmosphere and climate*, 2nd edn. Cambridge University Press, Cambridge, United Kingdom.

Semiletov, I. (1999) Aquatic sources and sinks of $CO_2$ and $CH_4$ in the polar regions. *Journal of the Atmospheric Sciences,* **56,** 286-306.

Takahashi, T., Olafsson, J., Goddard, J.G., Chipman, D.W. & Sutherland, S. (1993) Seasonal variation of $CO_2$ and nutrients in the high-latitude surface oceans: A comparative study. *Global Biogeochemical Cycles,* **7**, 843-878.

Takahashi, T., Sutherland, S.C., Wanninkhof, R., Sweeney, C., Feely, R.A., Chipman, D.W., Hales, B., Friederich, G., Chavez, F. & Sabine, C. (2009) Climatological mean and decadal change in surface ocean pCO2, and net sea–air CO2 flux over the global oceans. Deep Sea Research Part II: Topical Studies in Oceanography, 56, 554-577.

Talley, L.D. (2013) Closure of the global overturning circulation through the Indian, Pacific, and Southern Oceans: Schematics and transports, *Oceanography,* **2,** 80– 97.

Tison, J.-L., Delille, B. & Papadimitriou, S. (2017) Gases in sea ice. In: *Sea ice,* 3rd edn. (ed. by D.N. Thomas), pp. 433-471. Wiley, New Jersey.

Tremblay, J.-É., Anderson, L.G., Matrai, P., Coupel, P., Bélanger, S., Michel, C. & Reigstad, M. (2015) Global and regional drivers of nutrient supply, primary production and $CO_2$ drawdown in the changing Arctic Ocean. *Progress in Oceanography,* **139,** 171-196.

Vancoppenolle, M., Meiners, K.M., Michel, C., Bopp, L., Brabant, F., Carnat, G., Delille, B., Lannuzel, D., Madec, G. & Moreau, S. (2013) Role of sea ice in global biogeochemical cycles: emerging views and challenges. *Quaternary Science Reviews,* **79,** 207-230.

Vancoppenolle, M. & Tedesco, L. (2017) Numerical models of sea ice biogeochemistry. In: *Sea ice,* 3rd edn. (ed. by D.N. Thomas), pp. 492-515. Wiley, New Jersey.

Wiesenburg, D.A. & Guinasso Jr, N.L. (1979) Equilibrium solubilities of methane, carbon monoxide, and hydrogen in water and sea water. *Journal of Chemical and Engineering Data,* **24,** 356-360.

# Appendix 1

Methane rates at atmospheric CH$_4$ recording sites plotted against sea ice rate, selected to illustrate a range of phenologies.  Twin-peaked methane rates are typically plotted against the 'Arctic plus Antarctic' sea ice rate.  Sites are presented in ascending latitudinal order.  Each graph is provisional and should be checked before use, but we consider the similar features reliable.

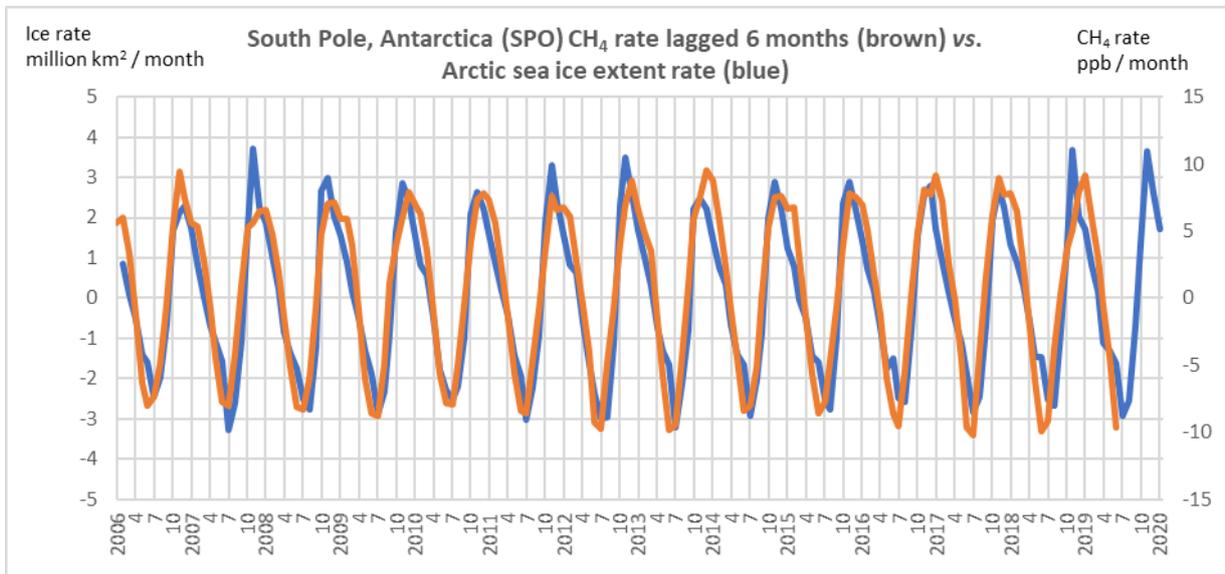

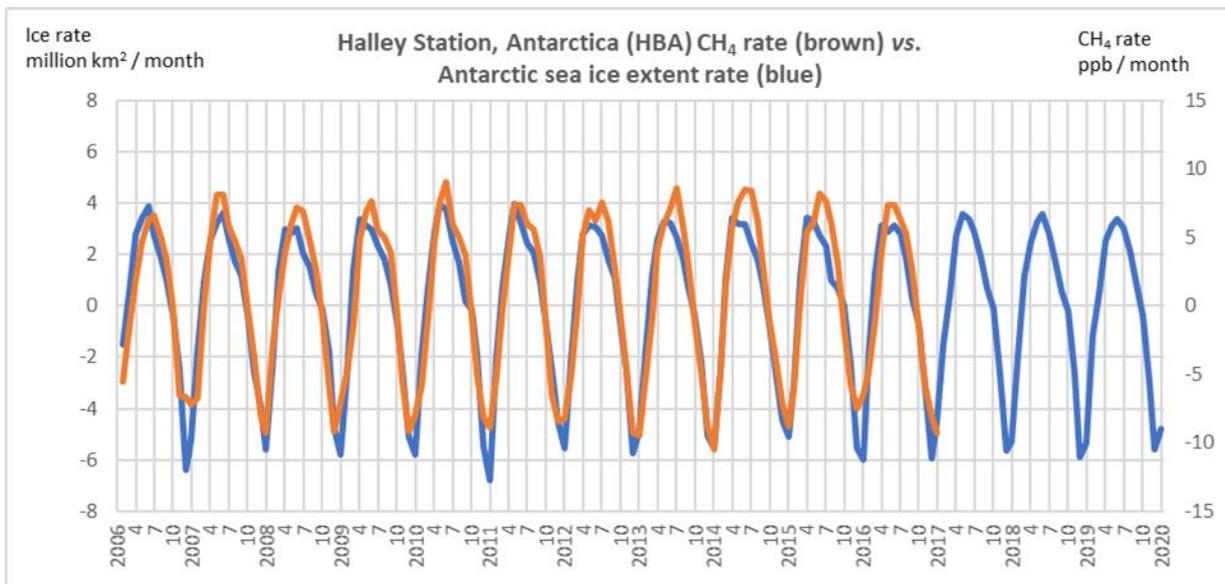

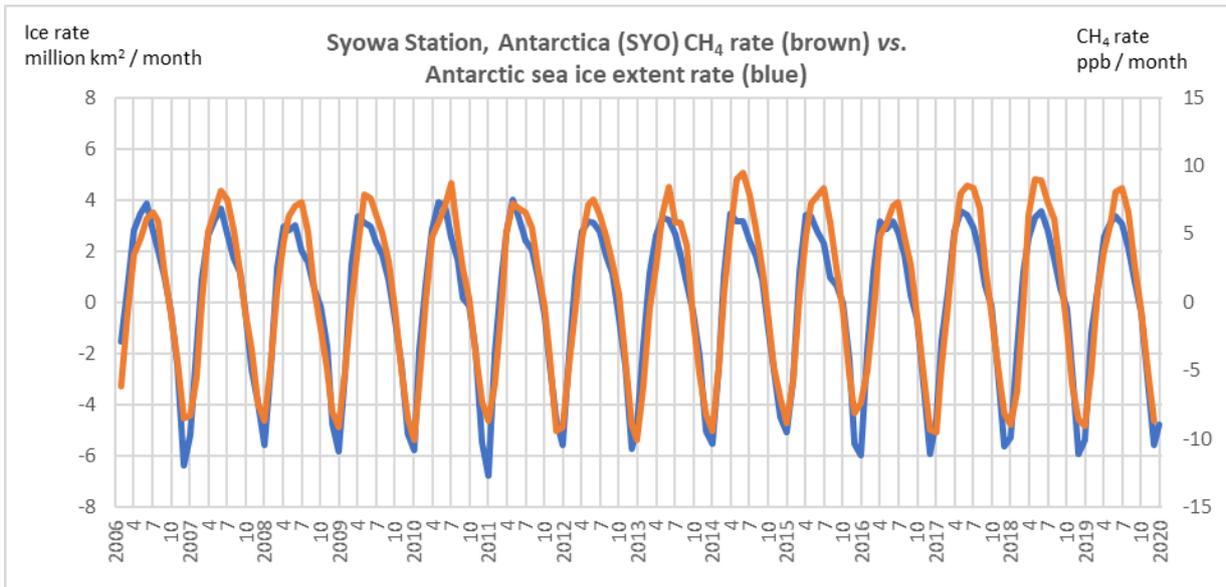

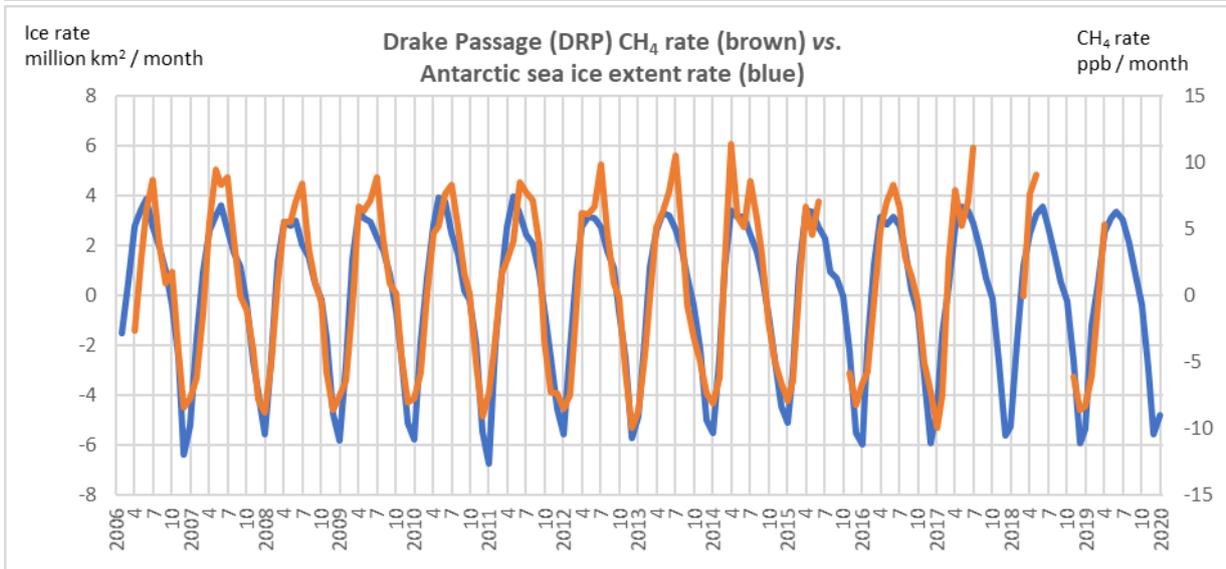

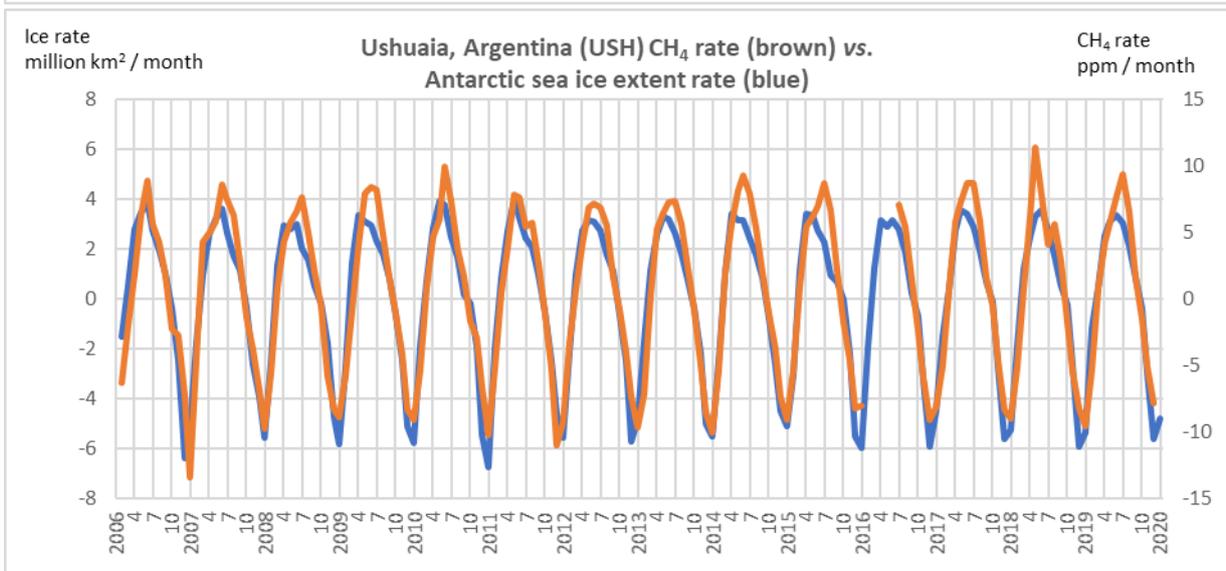

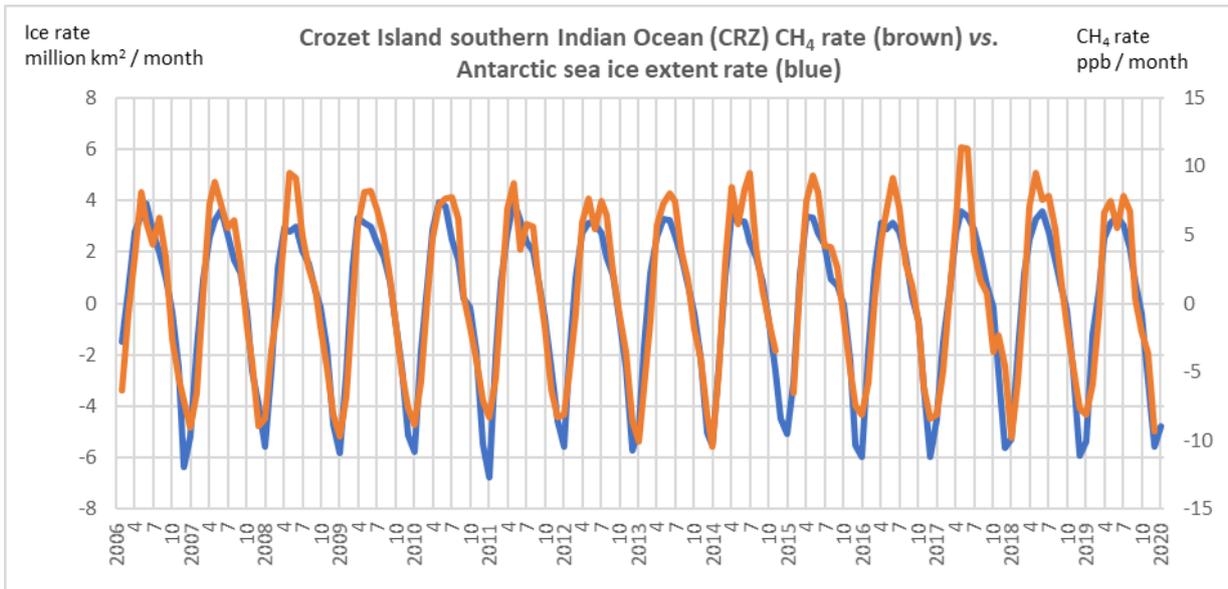

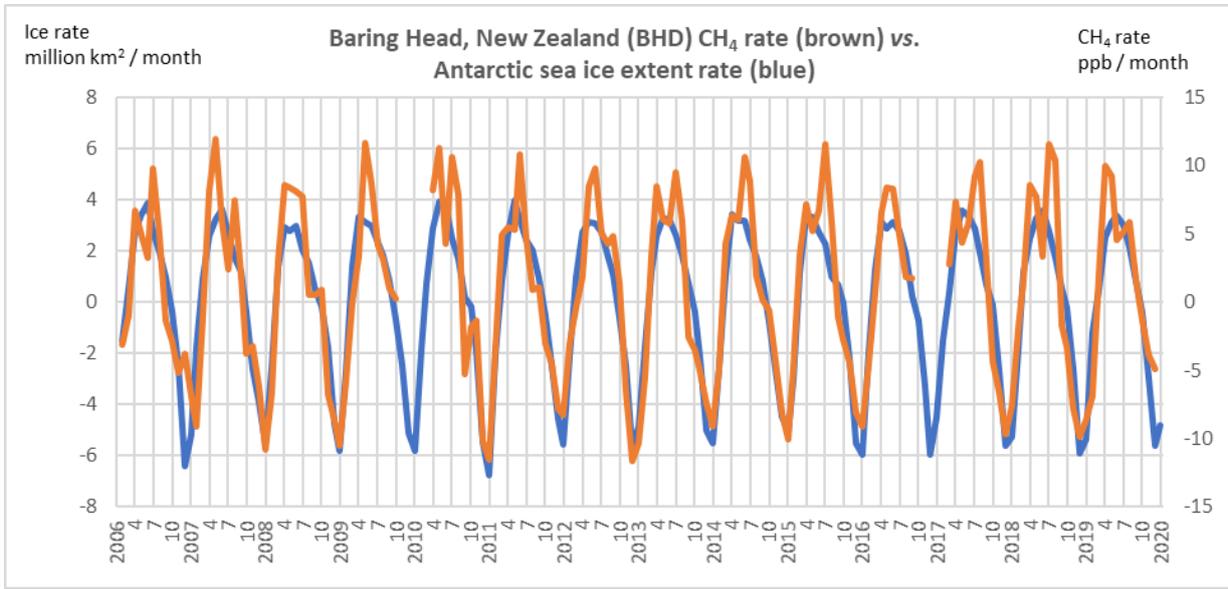

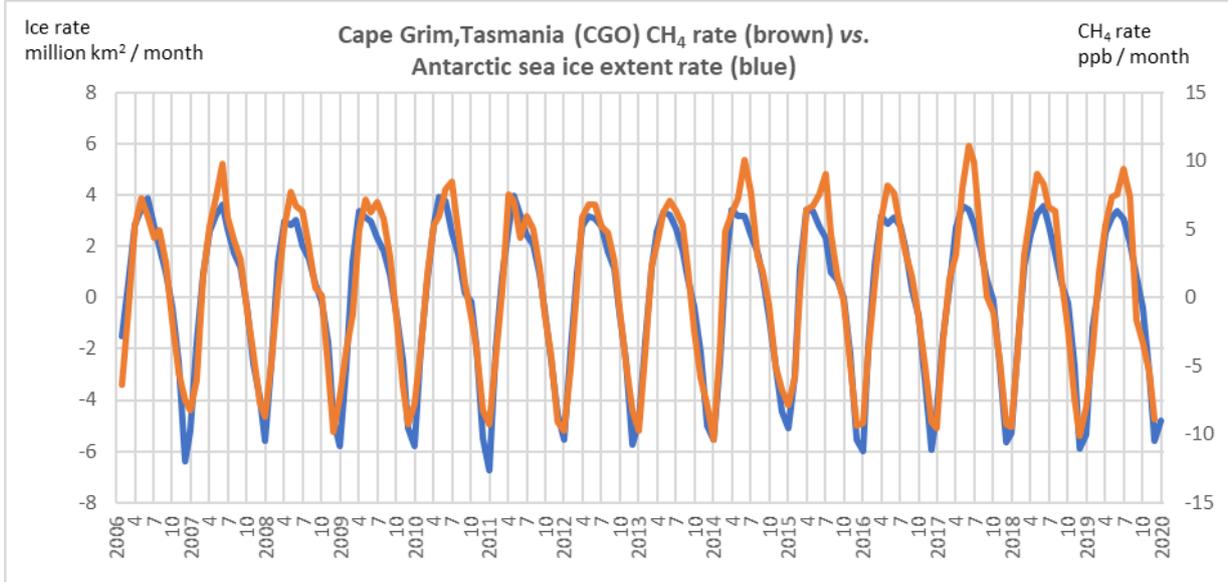

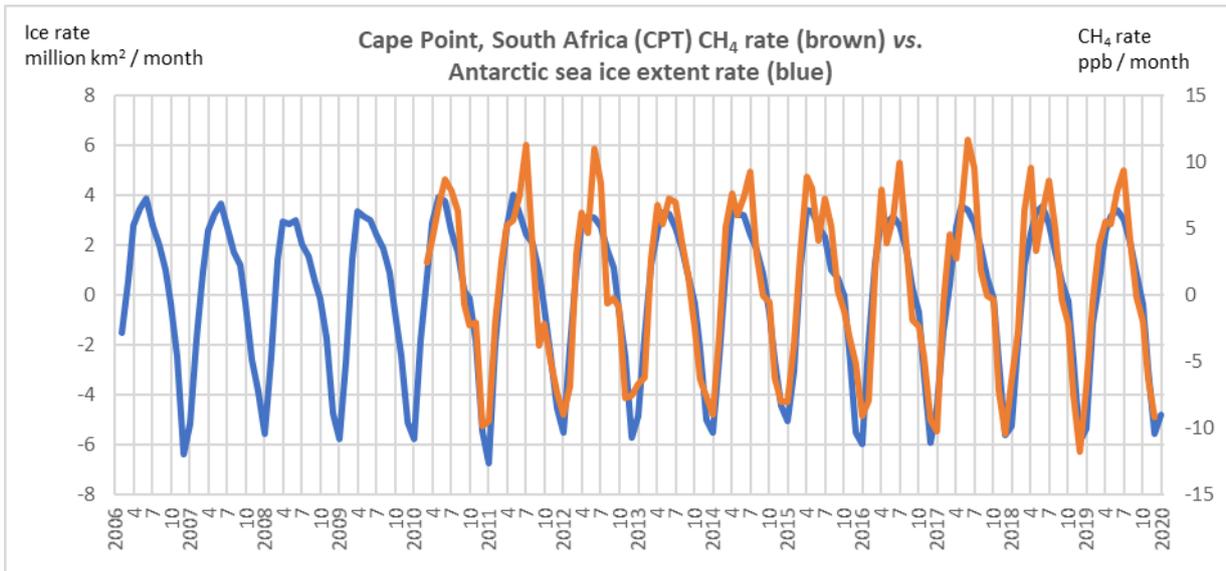

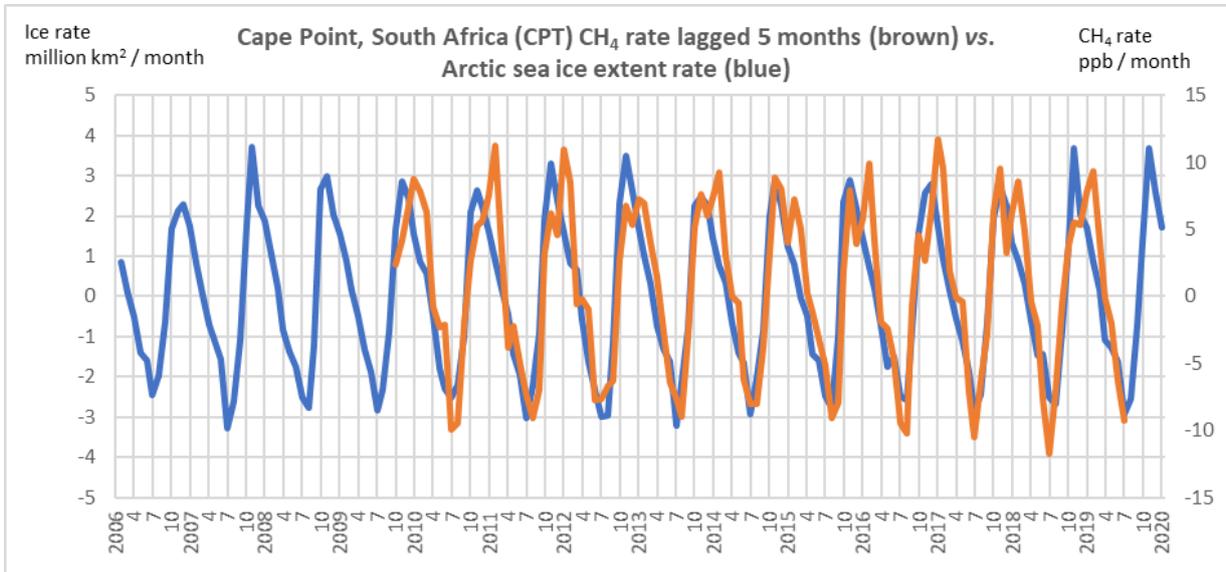

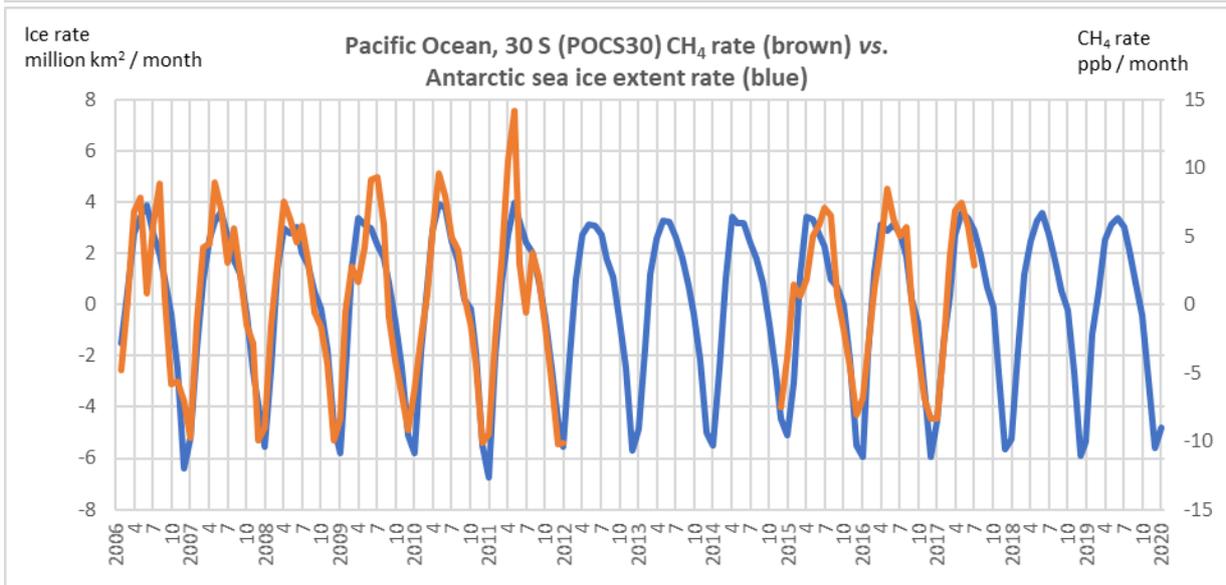

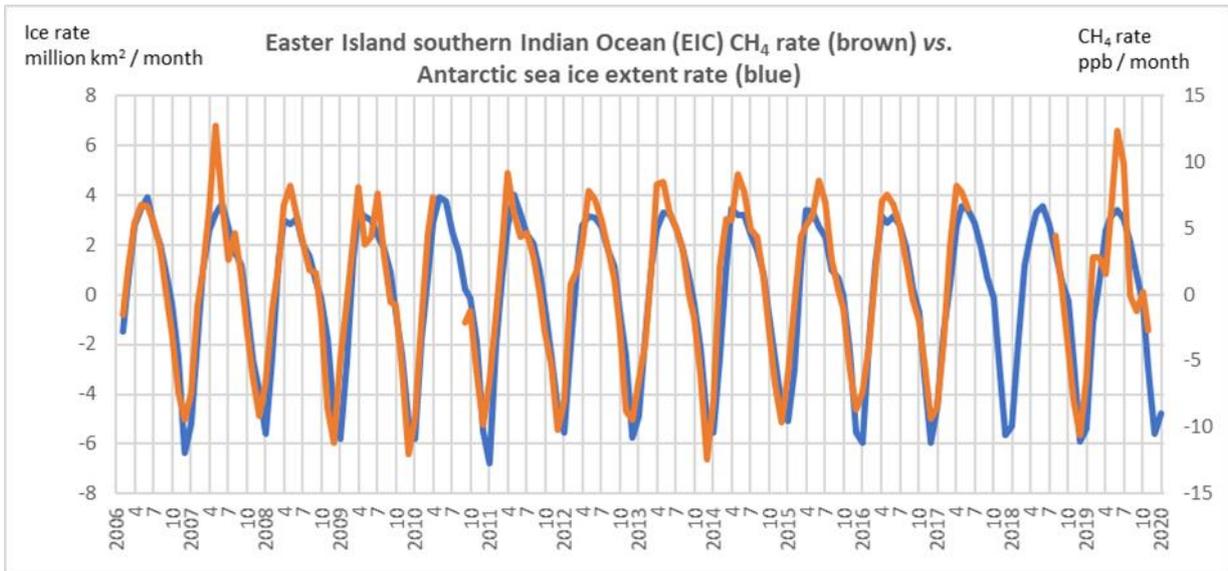

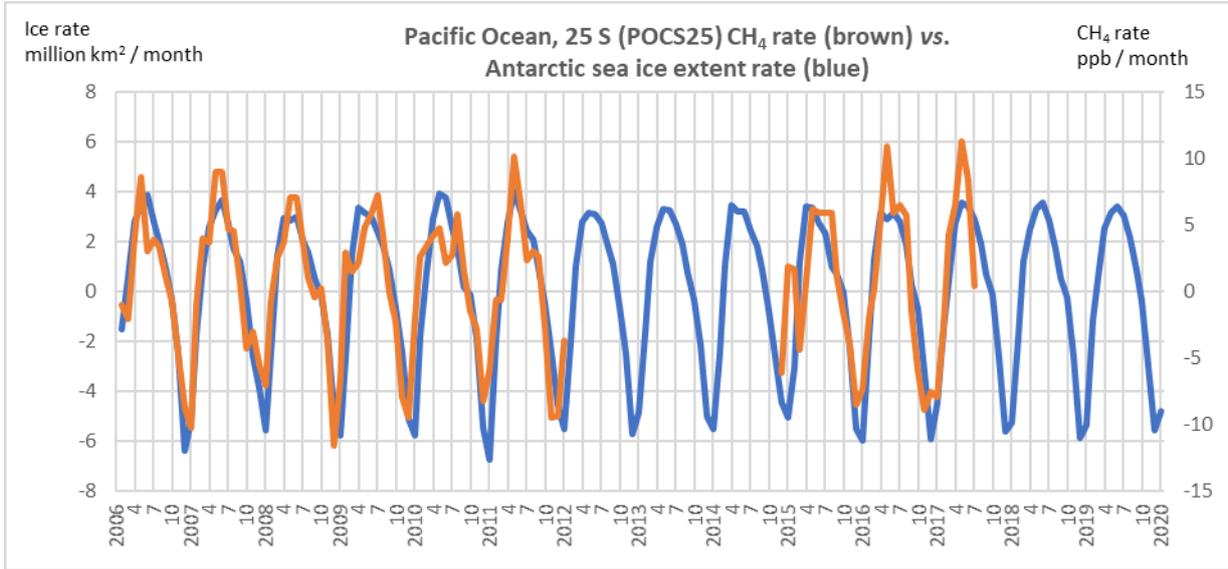

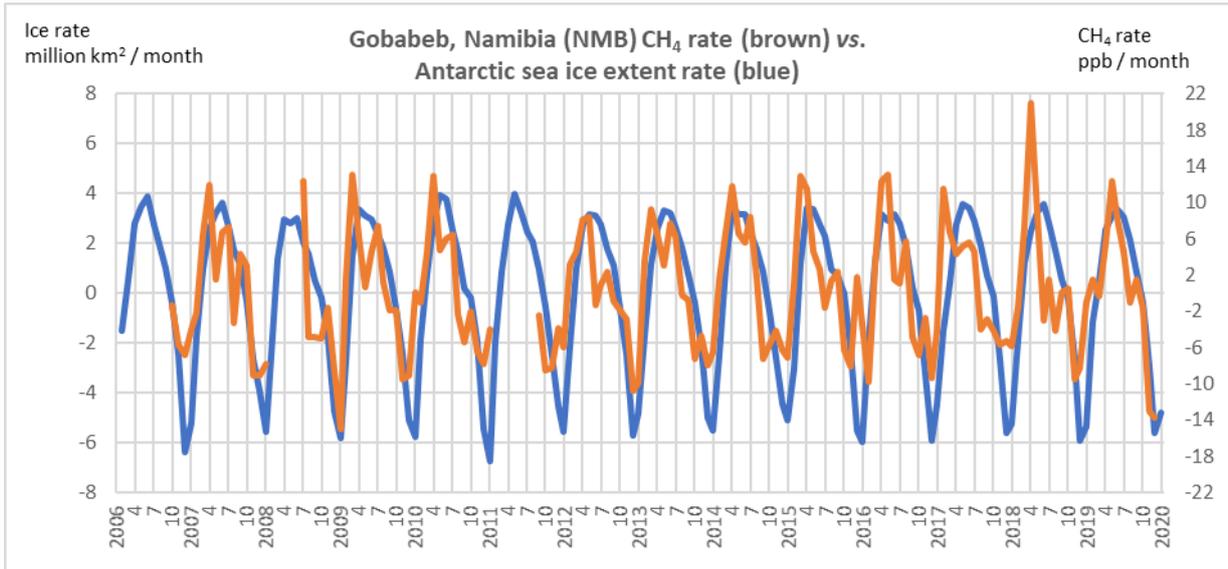

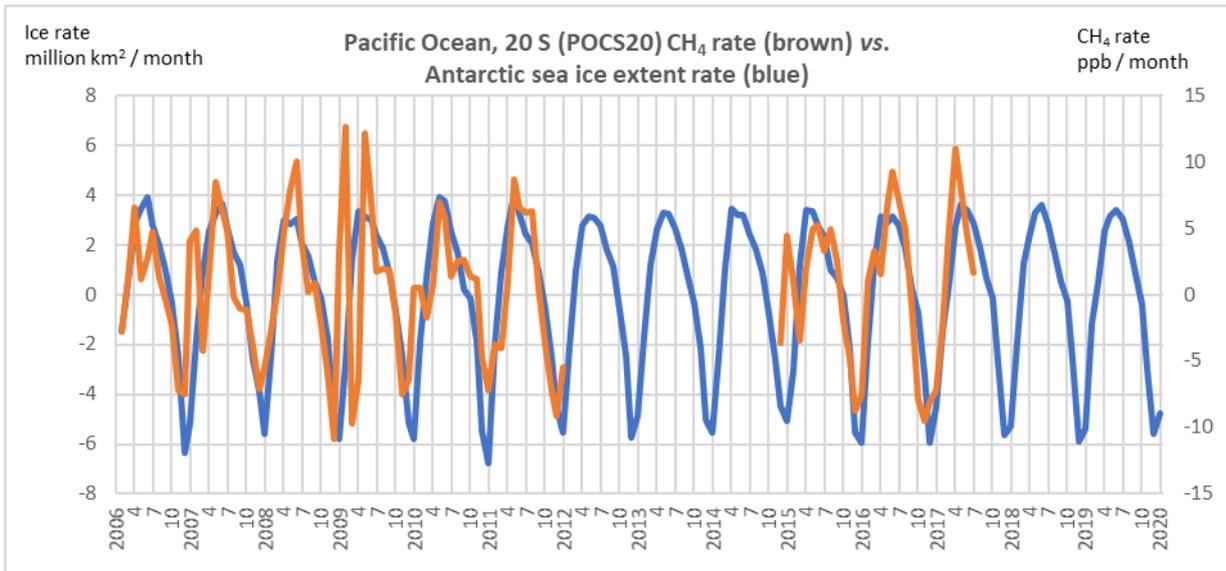

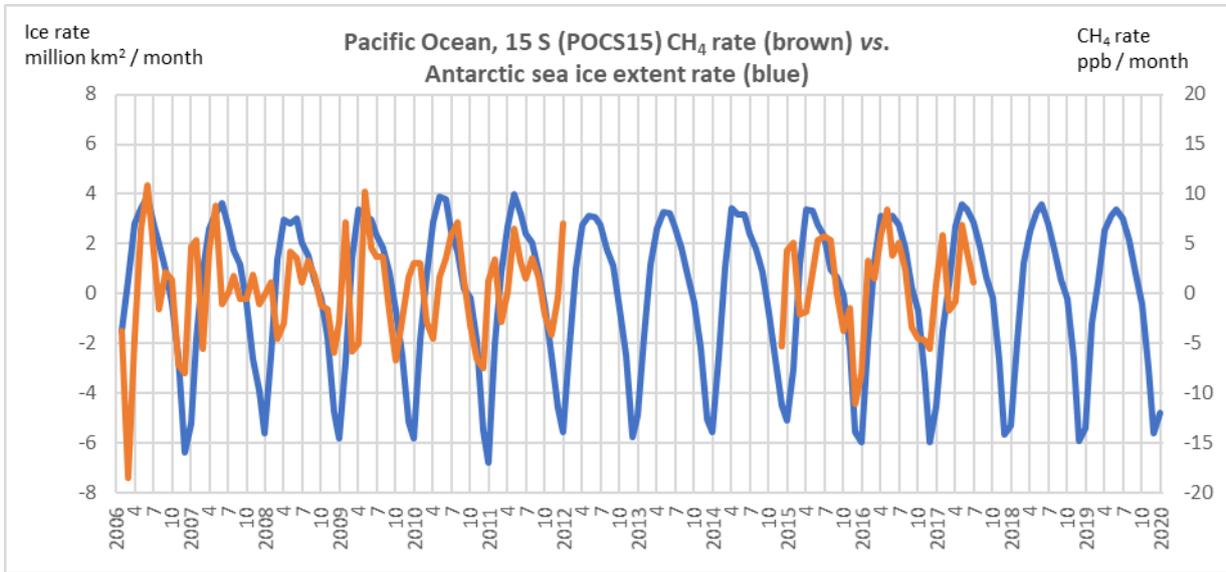

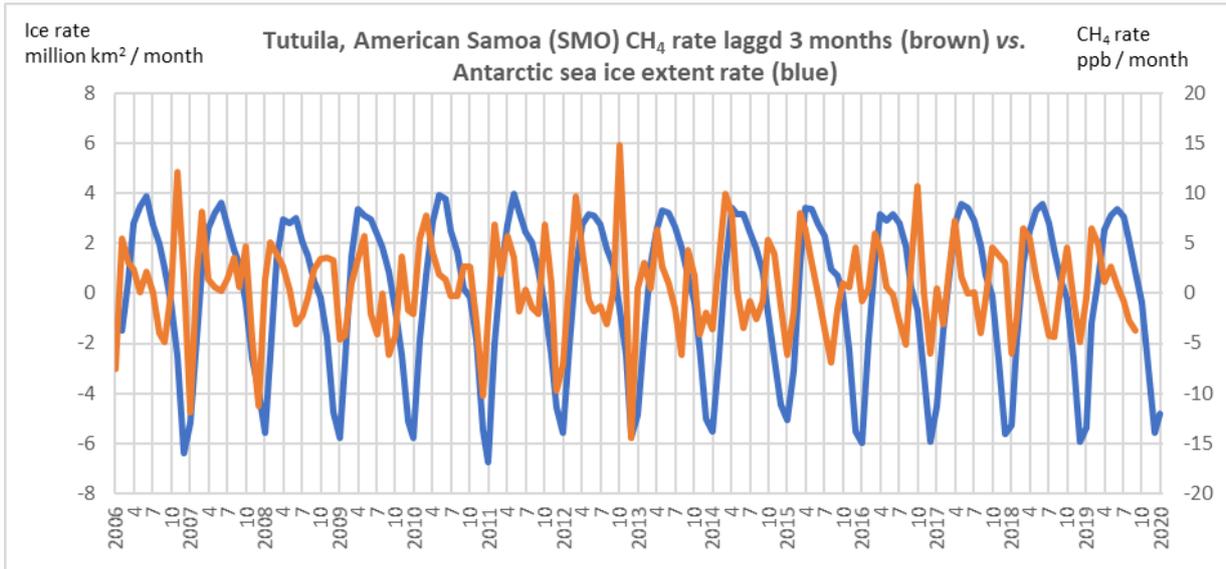

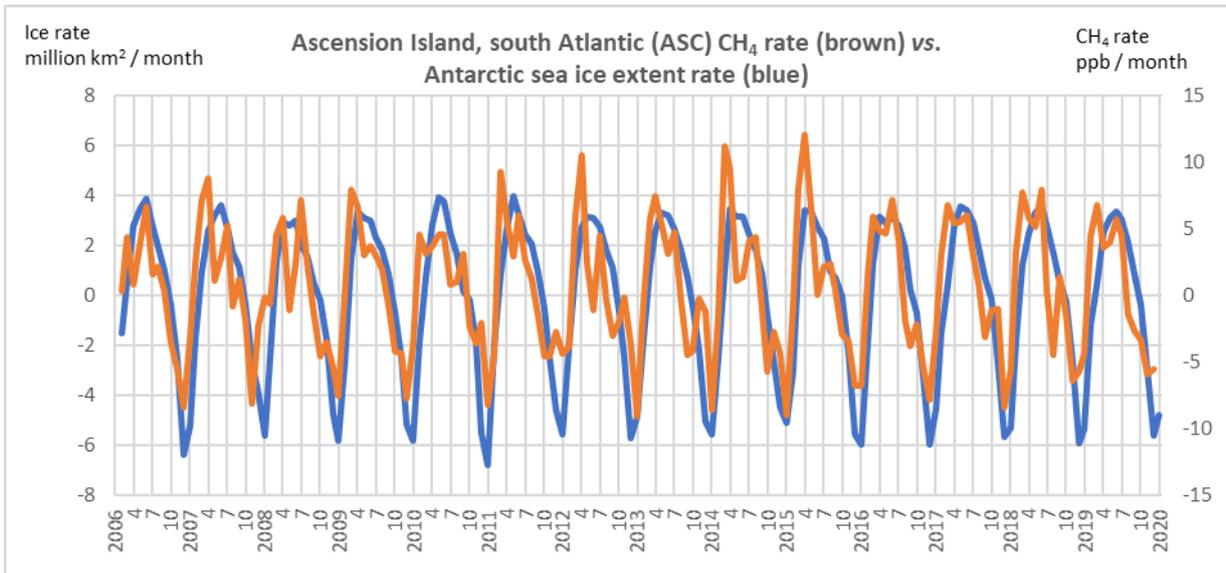

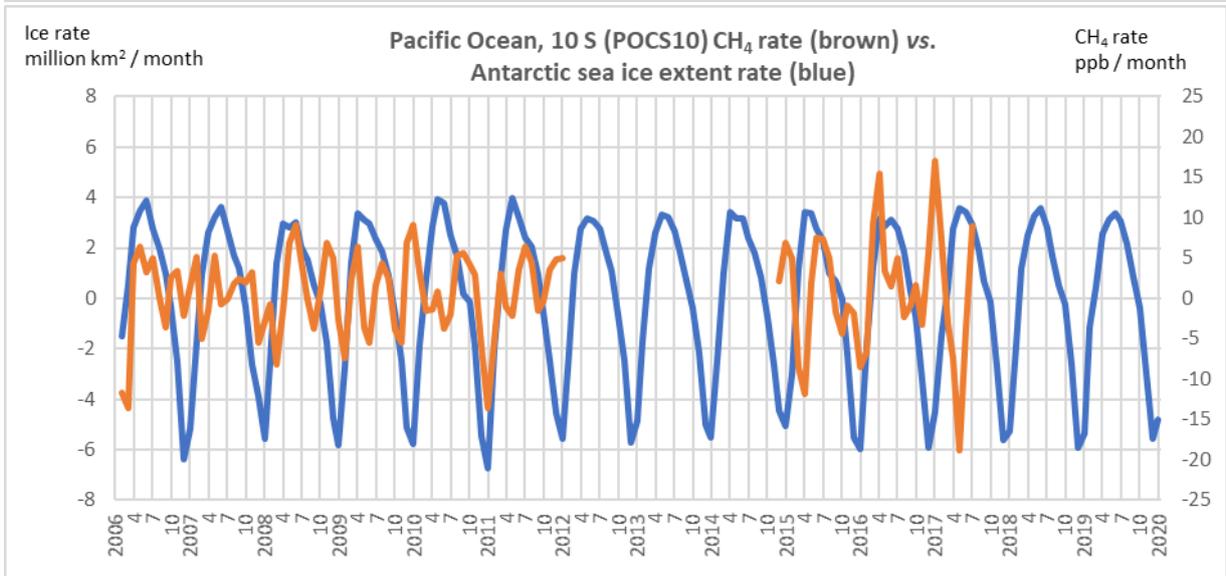

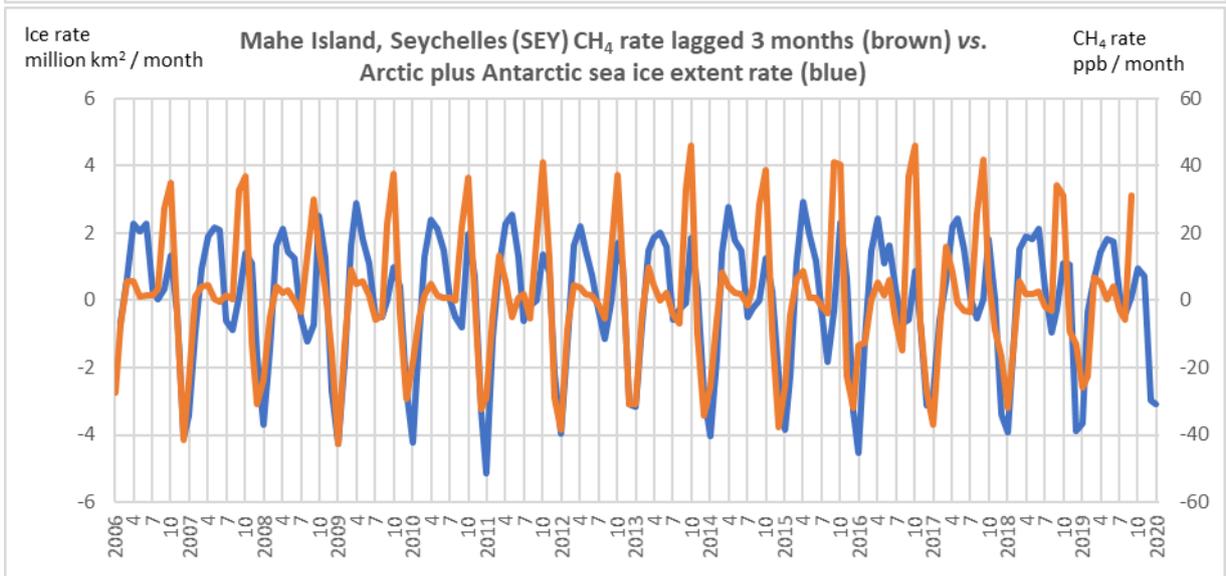

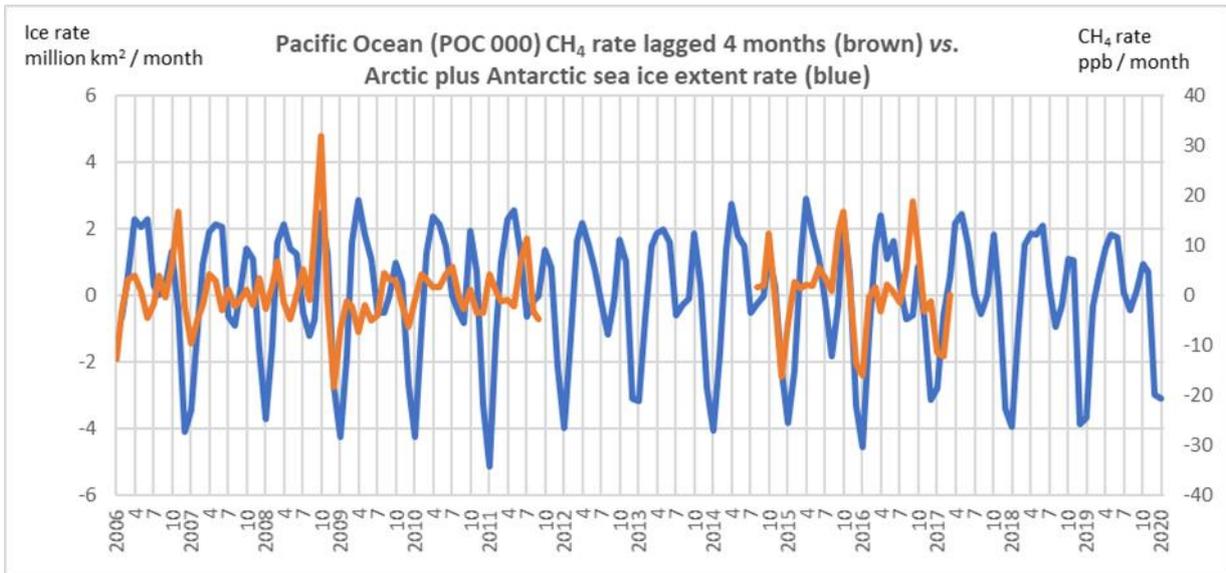

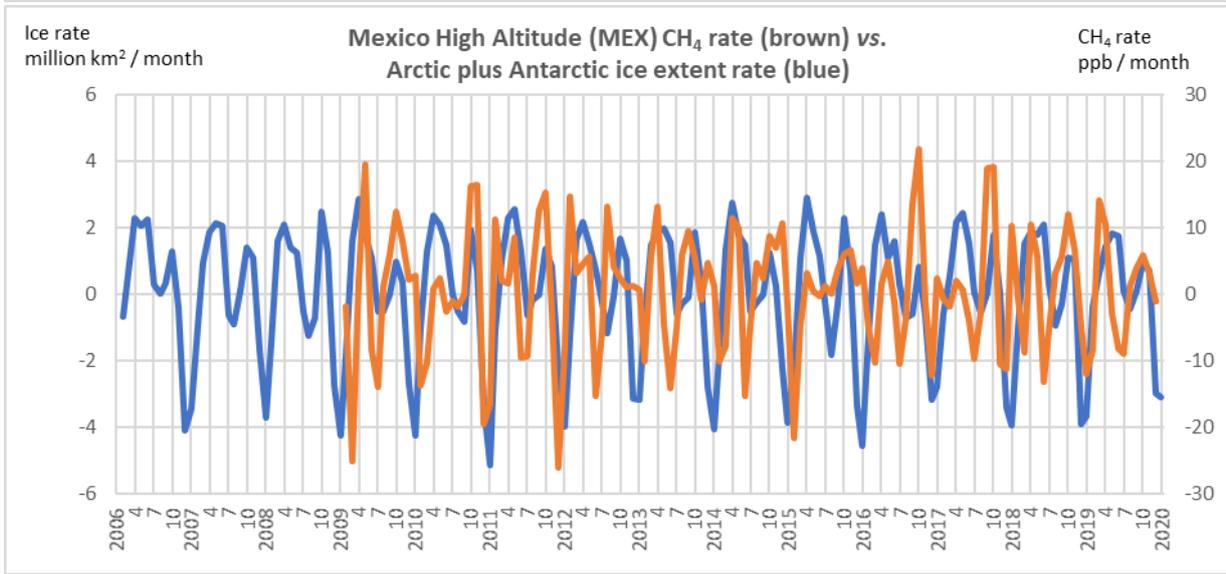

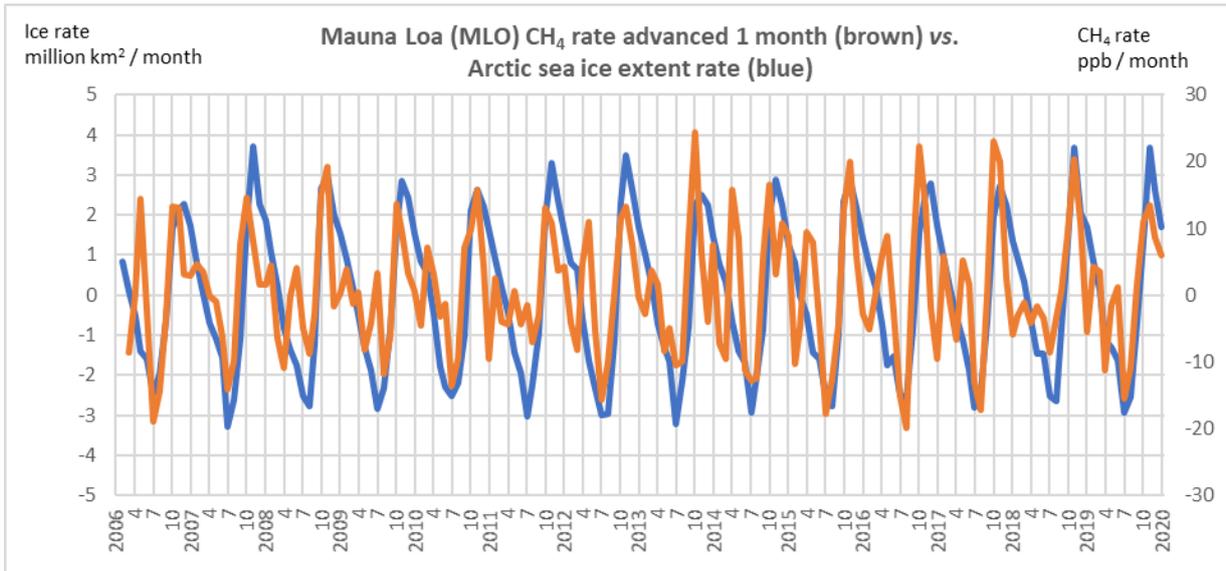

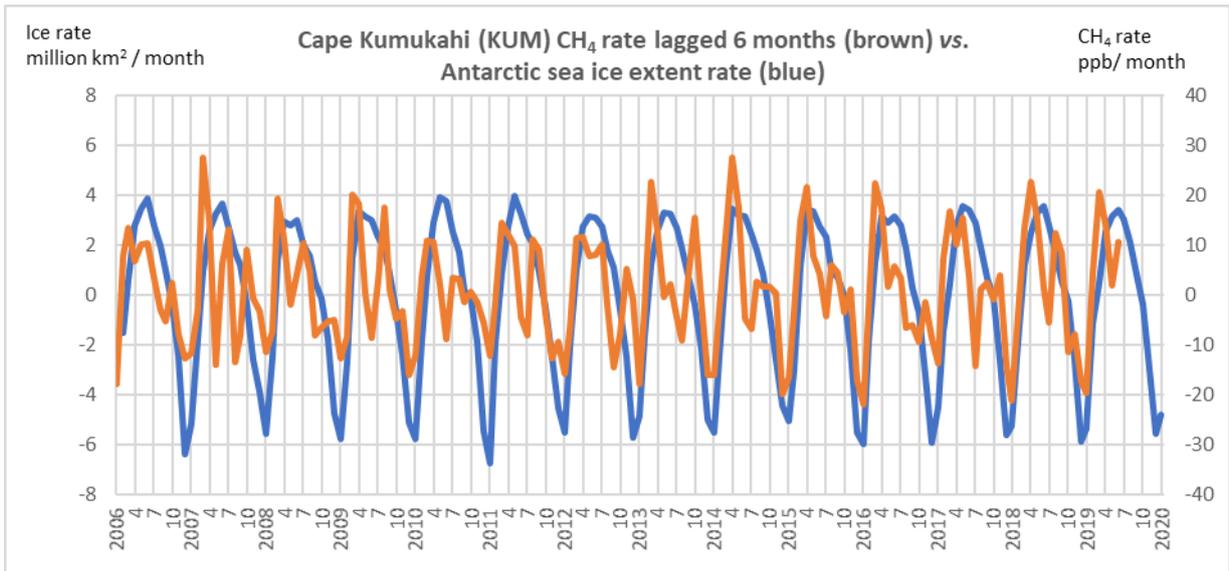

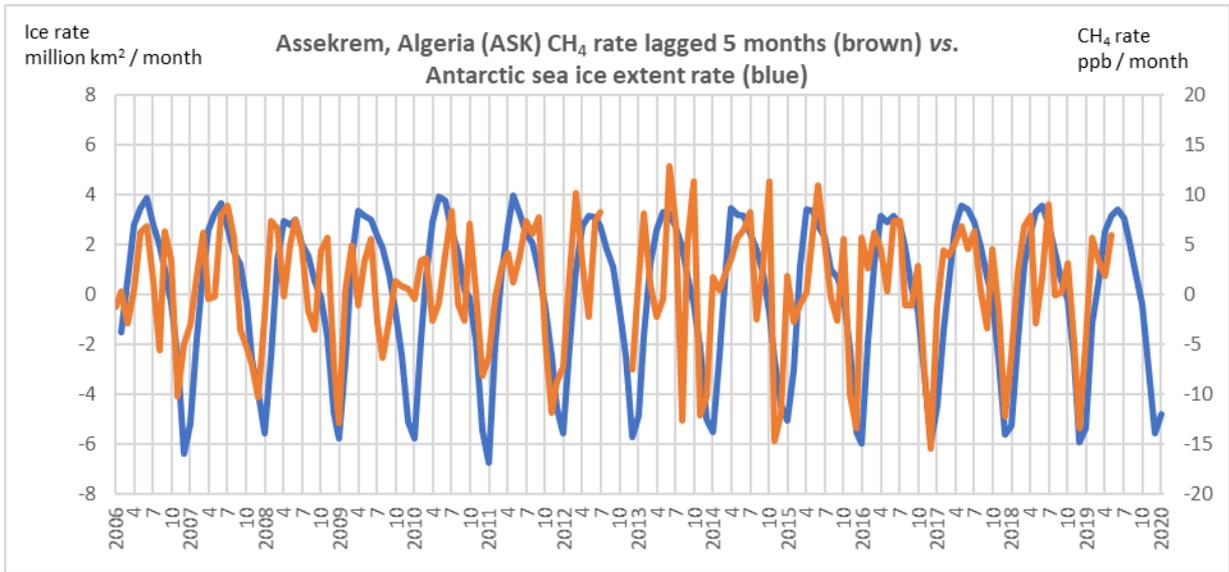

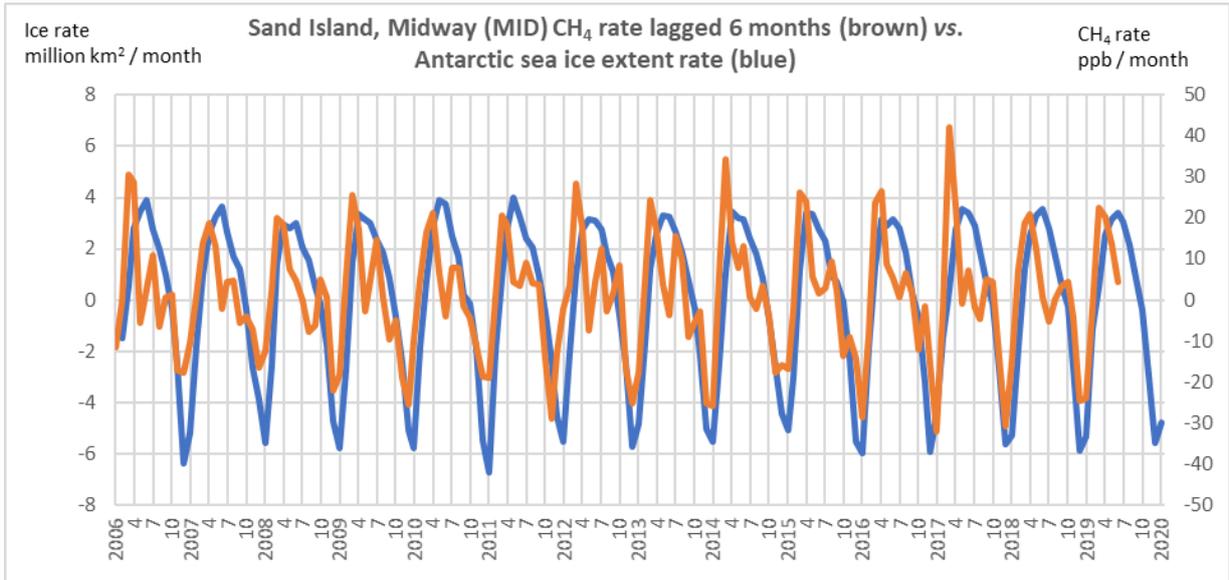

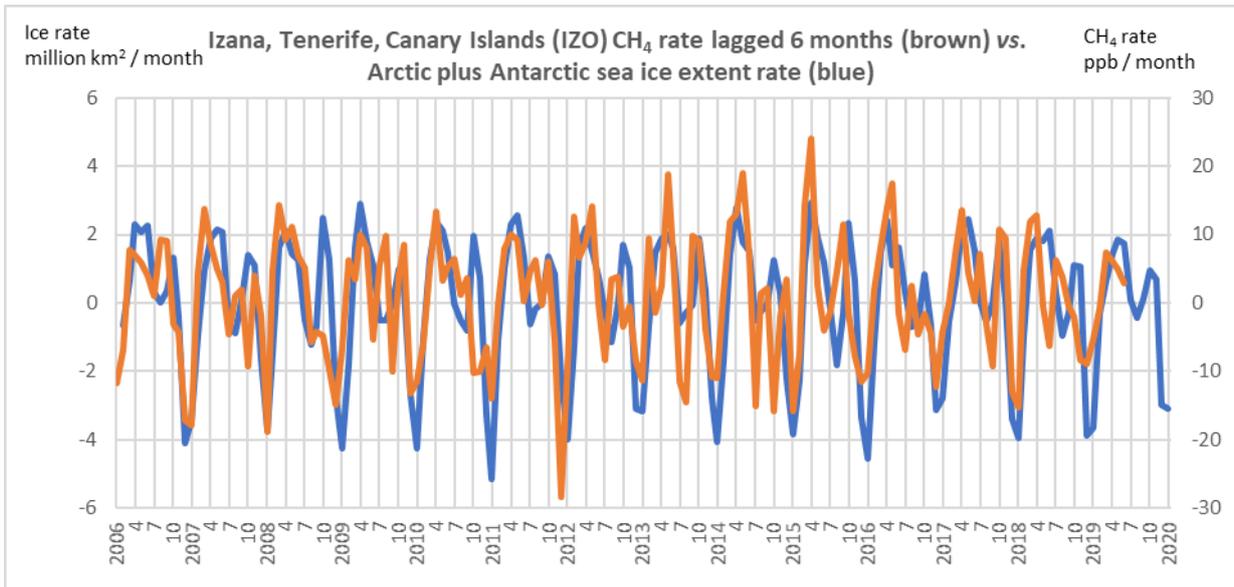

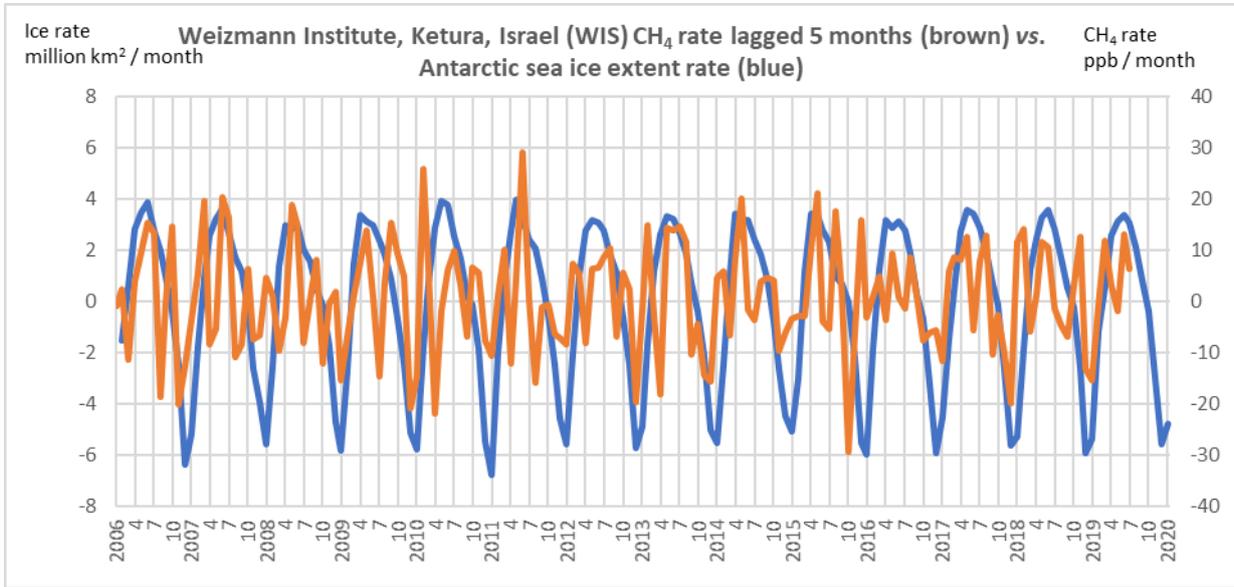

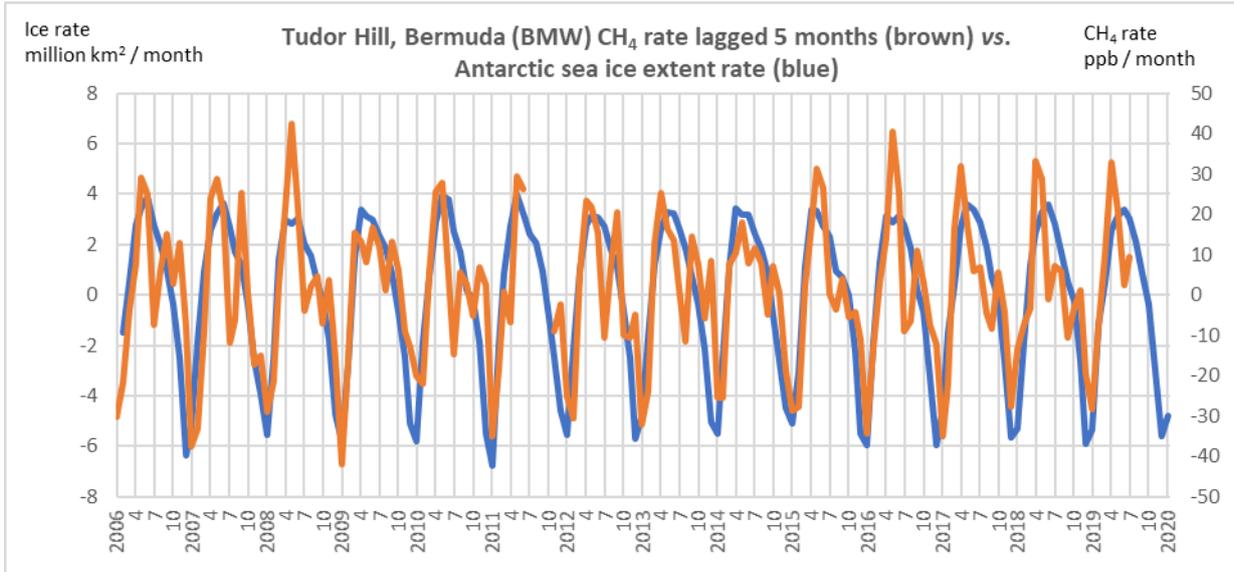

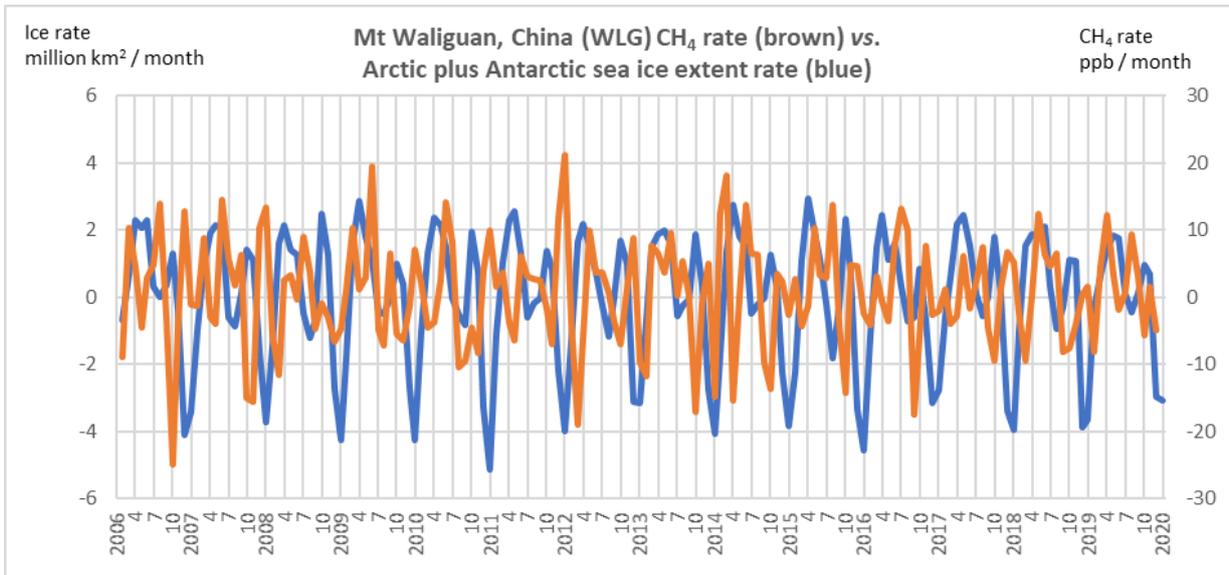

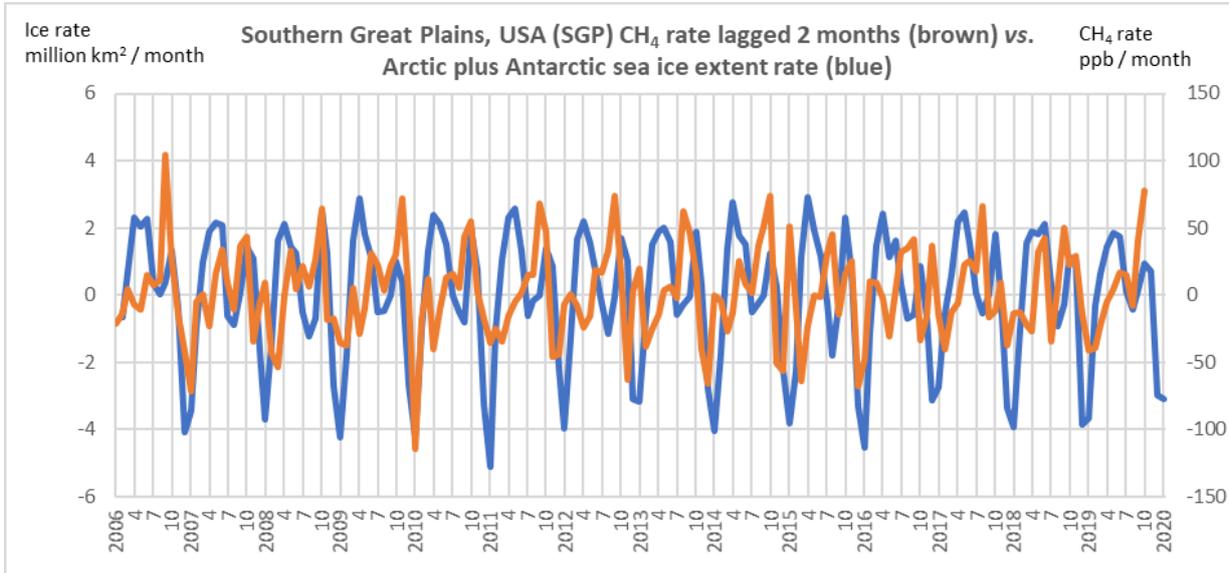

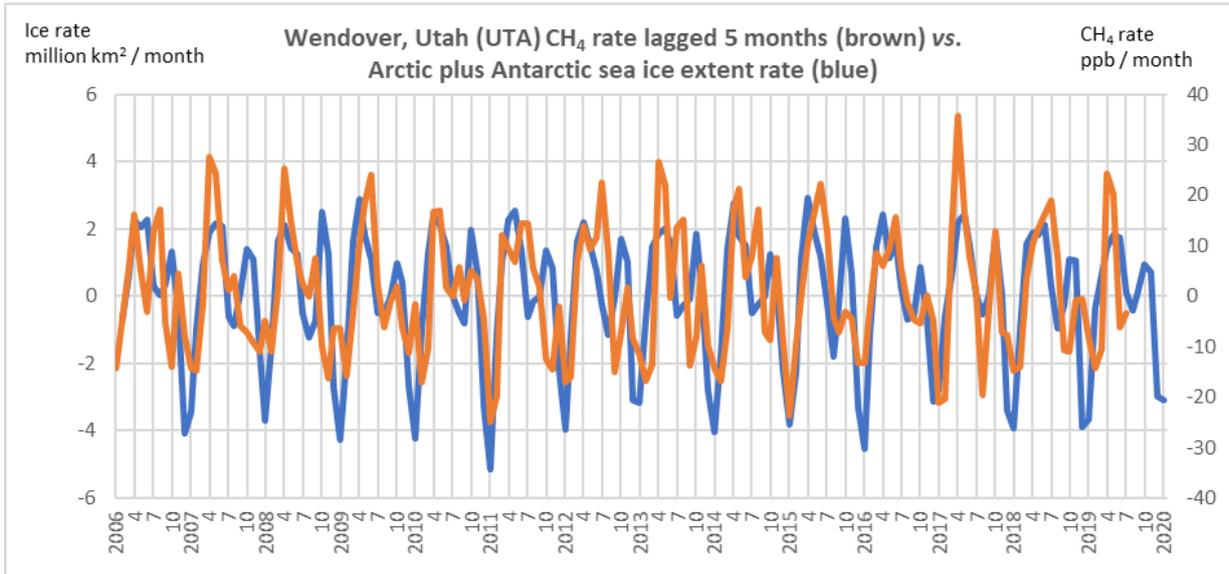

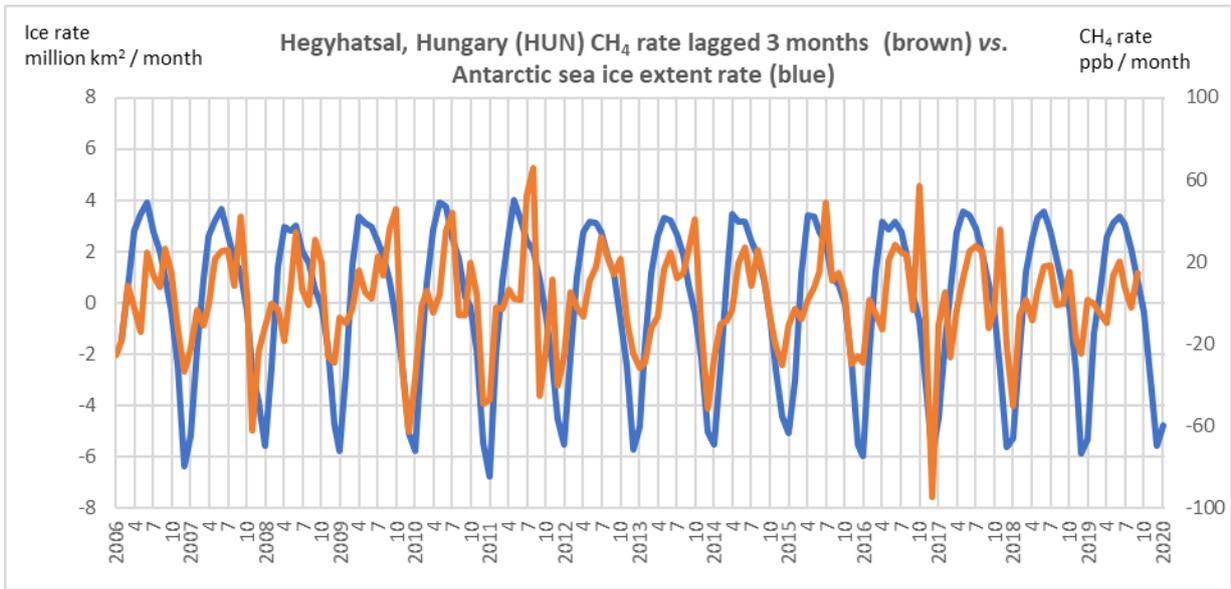

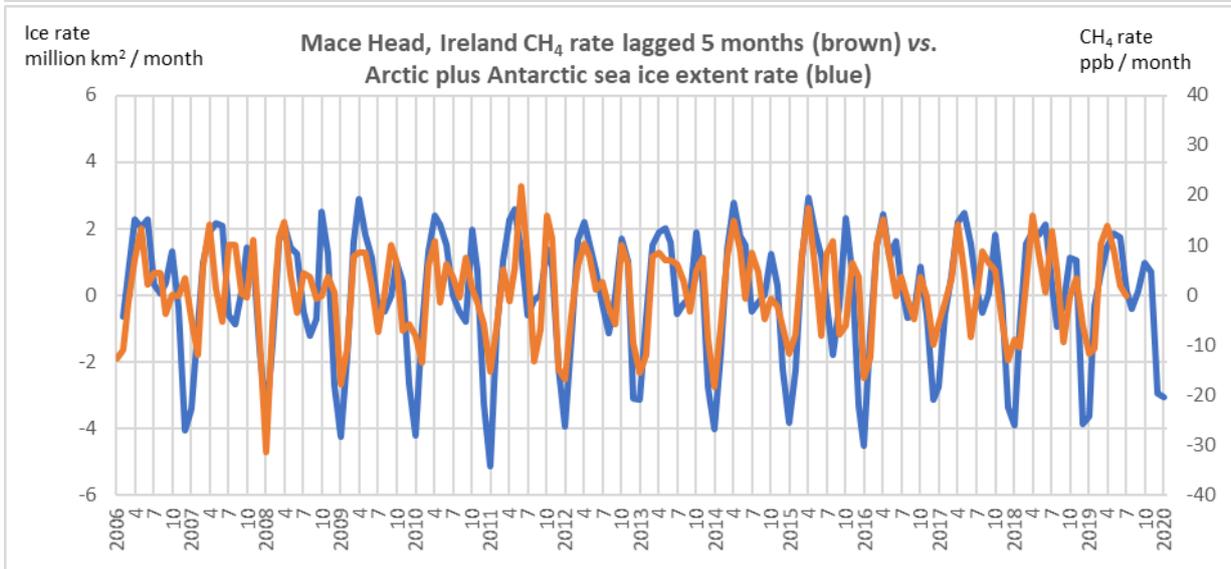

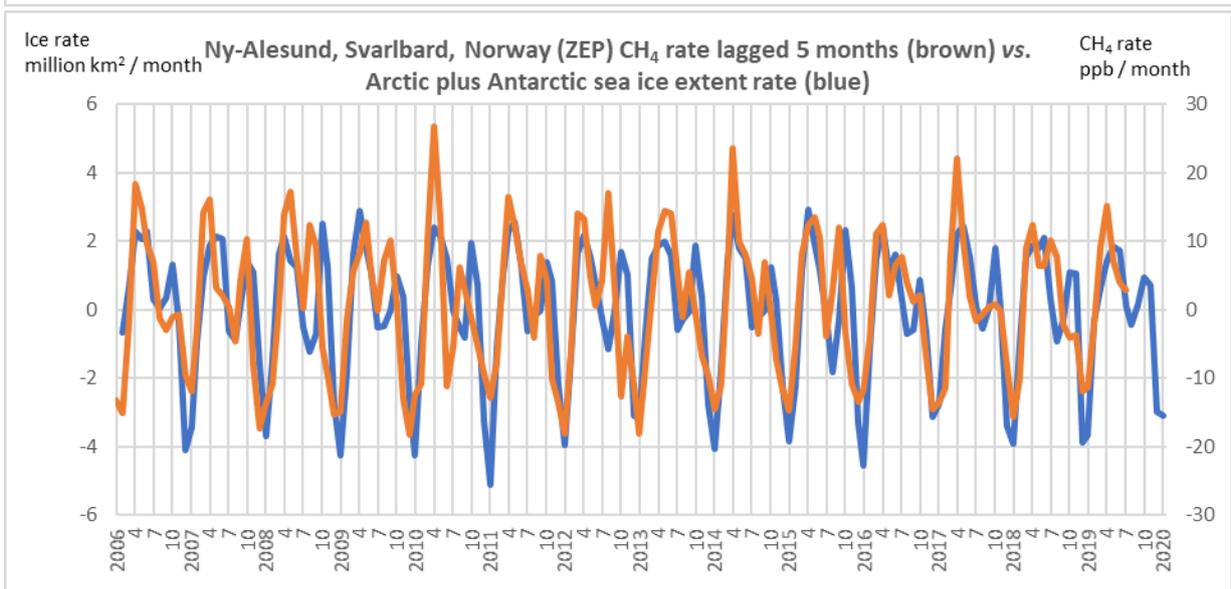

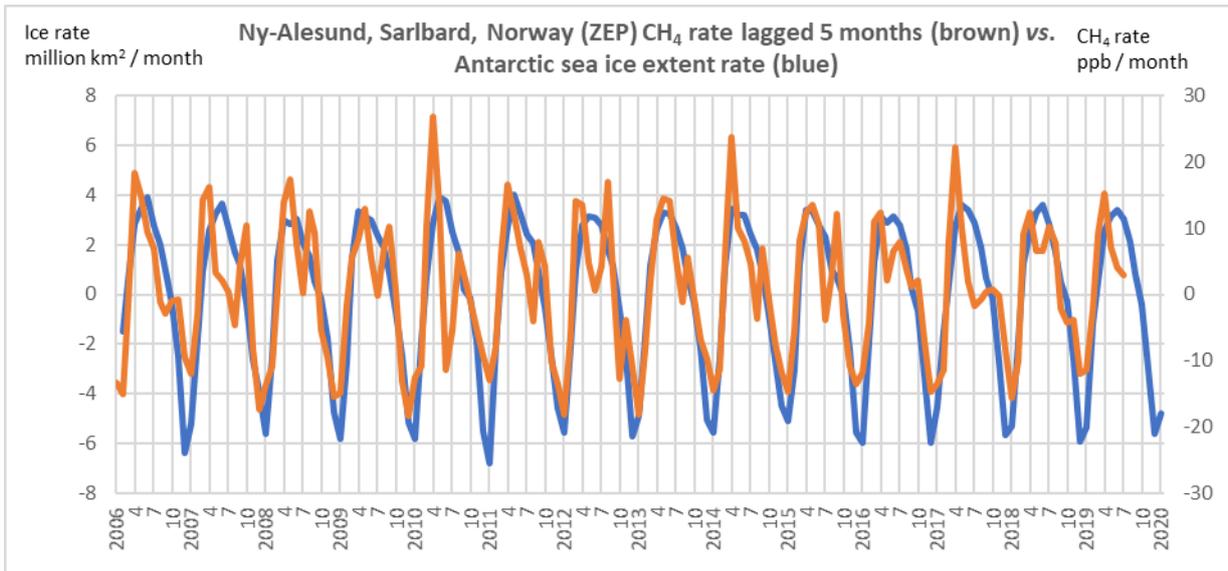

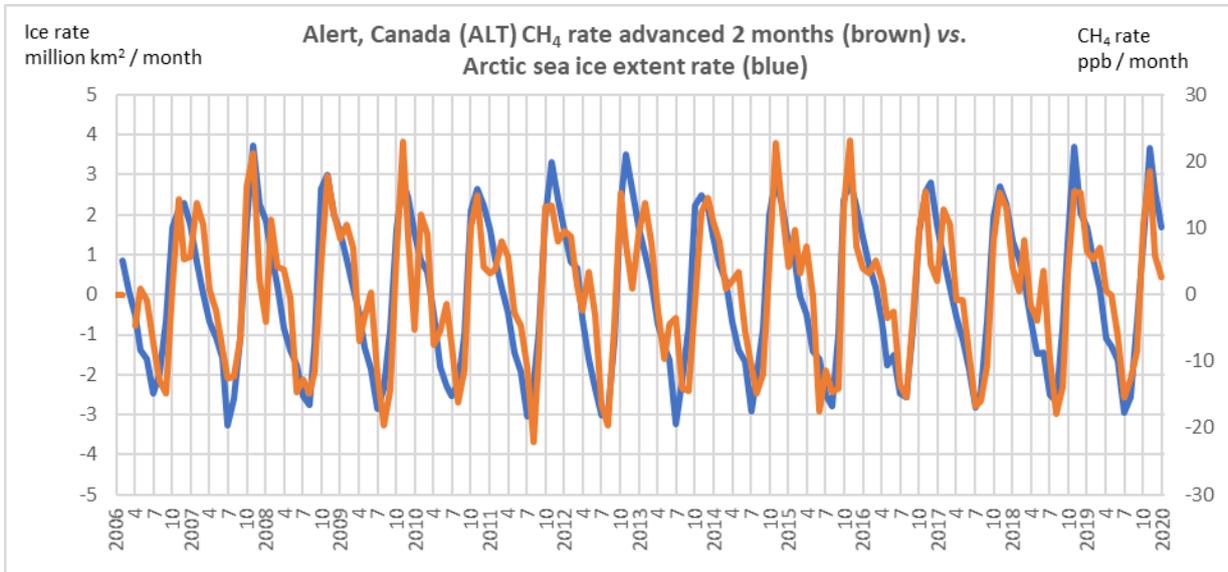

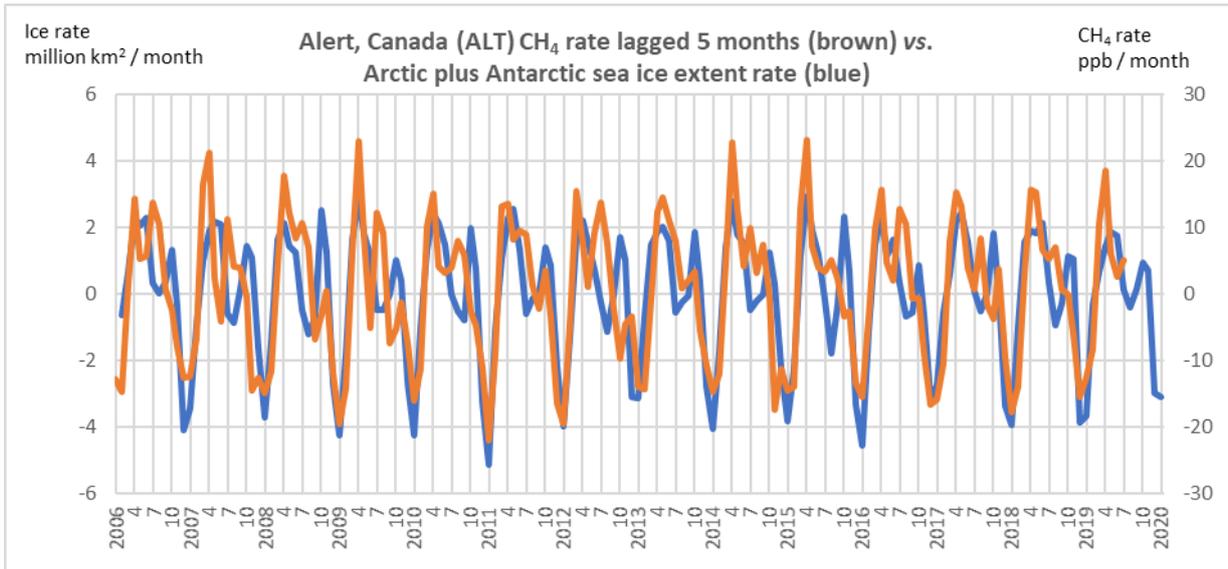

# Appendix 2

Carbon dioxide rate plotted against sea ice rate or methane rate, with lag to give a visual fit.  Carbon dioxide rates are calculated as for methane rates using sources in Table 1.

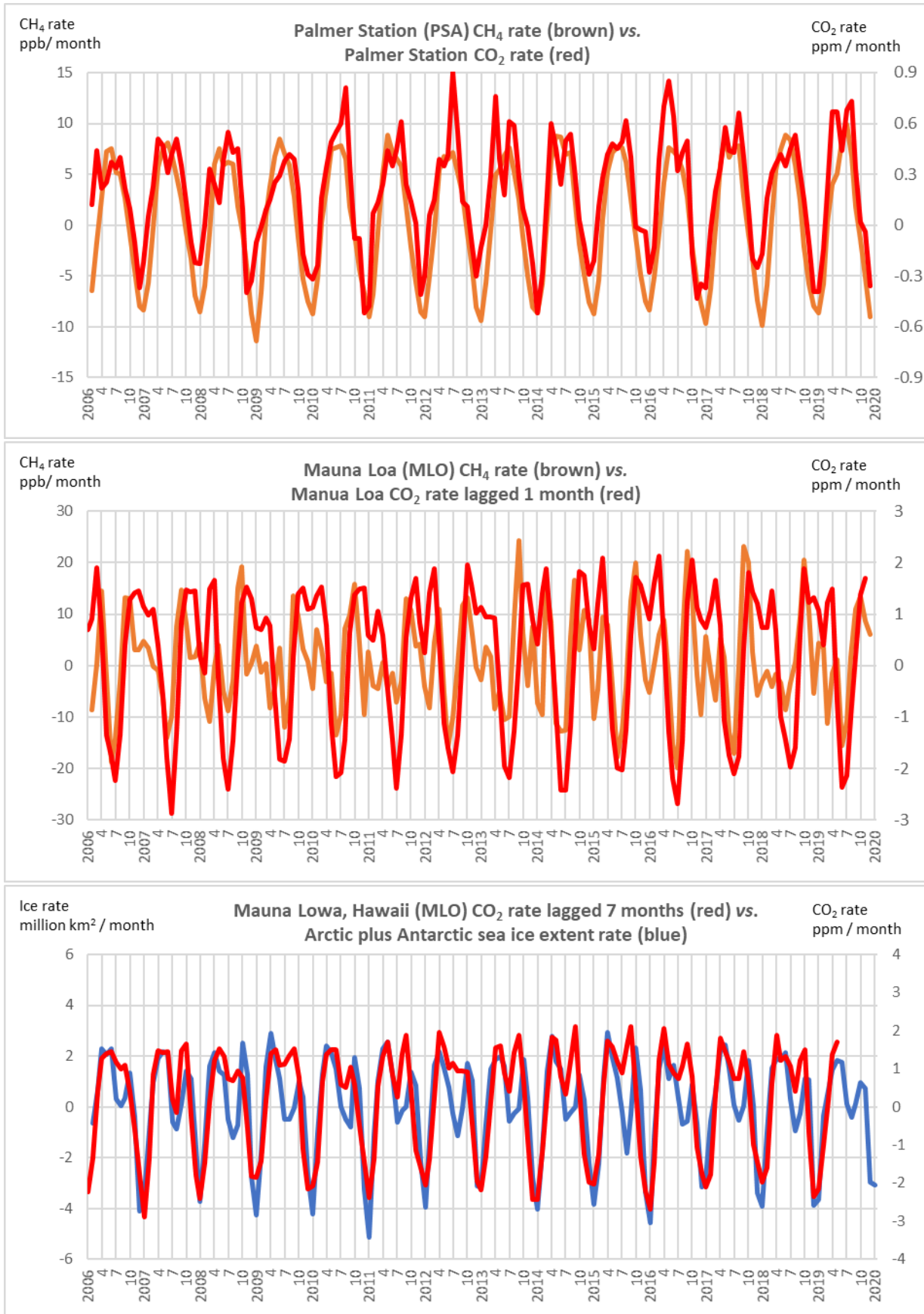